\documentclass[10pt]{article} 

\begin{document}

\newcommand{\vk}{{\vec k}}
\newcommand{\vK}{{\vec K}} 
\newcommand{\vb}{{\vec b}} 
\newcommand{{\vp}}{{\vec p}} 
\newcommand{{\vq}}{{\vec q}} 
\newcommand{\vQ}{{\vec Q}}
\newcommand{\vx}{{\vec x}}
\newcommand{\beq}{\begin{equation}}
\newcommand{\eeq}{\end{equation}} 
\newcommand{\half}{{\textstyle \frac{1}{2}}} 
\newcommand{\gton}{\stackrel{>}{\sim}}
\newcommand{\lton}{\mathrel{\lower.9ex \hbox{$\stackrel{\displaystyle
<}{\sim}$}}} \newcommand{\ee}{\end{equation}}
\newcommand{\ben}{\begin{enumerate}} 
\newcommand{\een}{\end{enumerate}}
\newcommand{\bit}{\begin{itemize}} 
\newcommand{\eit}{\end{itemize}}
\newcommand{\bc}{\begin{center}} 
\newcommand{\ec}{\end{center}}
\newcommand{\bea}{\begin{eqnarray}} 
\newcommand{\eea}{\end{eqnarray}}
\newcommand{\beqar}{\begin{eqnarray}} 
\newcommand{\eeqar}[1]{\label{#1} \end{eqnarray}} \newcommand{\pleft}{\stackrel{\leftarrow}{\partial}}
\newcommand{\pright}{\stackrel{\rightarrow}{\partial}}

\begin{center}
{\Large {\bf{ Heavy Quark Radiative Energy Loss in QCD Matter}}}

\vspace{1cm}

{ Magdalena Djordjevic and Miklos Gyulassy }

\vspace{.8cm}

{\em { Dept. Physics, Columbia University, 538 W 120-th Street,\\ New York,
       NY 10027, USA }} 

\vspace{.5cm}

\today
\end{center}

\vspace{.5cm}

\begin{abstract}
Heavy quark medium induced radiative energy loss 
is derived  to all orders in opacity, $(L/\lambda_g)^n$. The 
analytic expression generalizes the GLV opacity expansion for massless quanta
to heavy quarks with mass $M$ in a QCD plasma with a gluon dispersion
characterized by an asymptotic plasmon mass, $m_g=gT/\sqrt{2}$.
Remarkably, we find that the general result is obtained by 
simply shifting all frequencies in the GLV series 
by $(m_g^2+x^2 M^2)/(2 x E)$. Numerical evaluation of 
the first order in opacity energy loss shows
 that both charm and bottom energy losses
are much closer to the incoherent radiation limit
than light partons in nuclear collisions at both RHIC and LHC energies. 
However, the radiation lengths of heavy quarks
remain large compared to nuclear dimensions and hence high $p_T$
 heavy quark production is volume rather than surface dominated.
\end{abstract}

\section{Introduction}

The discovery~\cite{Adler:2003qi}-\cite{qm01} of a factor of $4\sim 5$
suppression of high $p_\perp \sim 5-10$ GeV hadrons produced in central $Au+Au$
at the Relativistic Heavy Ion Collider (RHIC) has been interpreted as evidence
for jet quenching of light quark and gluon jets~\cite{TOMO}-\cite{Wangsps}. Jet
quenching was predicted~\cite{Gyulassy:1990bh} to occur due to radiative energy
loss of high energy partons that propagate through ultra-dense QCD matter. The
observed quenching pattern is interesting not only as a test of QCD multiple
scattering theory but also because it provides a novel tomographic tool that
can be used to map the evolution of the quark gluon plasma plasma (QGP)
produced in ultra-relativistic nuclear collisions. Medium induced radiation
arises from higher twist final state interactions and depends on the integrated
optical thickness or gluonic opacity $L/\lambda_g$ 
of the QCD medium.  At high $p_T$ it dominates
over the energy loss due to elastic scattering~\cite{Bjorken:1982tu}.

Recent data from RHIC $D+Au$ reactions~\cite{Adler:2003ii}-\cite{Back:2003ns},
on the other hand, show the absence of jet quenching in such light ion beam 
reactions. These data are in accord with predictions based on jet tomography 
in Refs. \cite{Gyulassy:2003mc,Wang:1998ww,Vitev:2002pf} because in light ion
interactions no extended dense QCD medium is produced in the final state.
This important $D+Au$ control experiment therefore strengthens
 the interpretation
of the observed high $p_T$ quenching pattern in central $Au+Au$ as 
due to final state energy loss of 
jets  produced in dense QGP matter.

The first estimates for heavy quark energy
loss~\cite{Shuryak:1996gc}-\cite{Lin:1998bd} proposed that similar quenching
may occur for charm jets as for light partons. However,
in ref.~\cite{Dokshitzer:2001zm} it was pointed out 
that the heavy quark mass leads
to a kinematical ``dead cone'' effect for $\theta<M/E$ that reduces
significantly the induced radiative energy loss of heavy quarks.  Numerical
estimates indicated that the quenching of charm quarks may be 
about a half that of light quarks.  Experimentally, the first PHENIX
data~\cite{Adcox:2002cg} on ``prompt'' single electron production in $Au + Au$
collisions at $\sqrt{s}=130$ AGeV provided a first rough look at heavy quark
transverse momentum distributions at RHIC.  Remarkably, no indication for a QCD
medium effect was found within the admittedly large experimental errors (see
also ref.~\cite{Batsouli_Gyulassy}). However, in the near future, data with
much higher statistics and wider $p_T$ range will become accessible.
 
We concentrate in this paper on the theory of heavy quark energy loss,
extending our previous works I~~\cite{Djordjevic:2003qk} and 
II~\cite{Djordjevic:2003be}.
Their quenching pattern and correlations should 
provide independent complementary
tests of QGP production.  In addition, heavy quark observables can be used to
test different approximations in the emerging theories of heavy quark radiative
energy loss QCD~\cite{Shuryak:1996gc}-\cite{Dokshitzer:2001zm}
,~\cite{Djordjevic:2003qk}-\cite{Zhang:2003wk}.  The main new
feature that heavy quark probes introduce is a controlled reduction of the
radiated gluon formation times due to finite mass kinematic effects.  As the
mass increases. the increase of phase shifts leads to a reduction~\cite{Djordjevic:2003qk}
of destructive Landau-Pomeanchuk-Migdal (LPM) interference 
effects 
that were found 
to be so important for light quark or gluon 
jet energy loss~\cite{TOMO}-\cite{Wangsps}. 
In addition, the kinematic reduction of propagator amplitudes
leads to an overall decrease of  the magnitude of heavy energy loss
relative to light jet energy loss 
even in the incoherent Gunion-Bertsch~\cite{Bertsch-Gunion} (GB) limit, 
as we demonstrate quantitatively below. 
Therefore,  the predictable high $p_T$ 
open charm and open bottom  observables at RHIC and LHC
will provide interesting new control tests of 
QCD dynamics in nuclear collisions.

The main goal of this paper is (1) to generalize the GLV opacity 
series~\cite{GLV} to include massive quark kinematic effects and (2) to take 
into account the Ter-Mikayelian plasmon effects for gluons as described in 
Ref.~\cite{Djordjevic:2003be}. The competition between these two medium 
effects was first discussed in our paper I~\cite{Djordjevic:2003qk}. We 
showed that the apparent null effect observed for heavy quarks via single 
electrons could in part be due to a reduction of the leading $O(\chi^0)$ 
order radiation {\em associated} with the initial hard production process.
In this sense, there are two opposing medium effects: (1) at $O(\chi^0)$, the
Ter-Mikayelian plasmon dispersion effect that reduces the {\em associated} 
hard radiation energy loss~\cite{Djordjevic:2003be} and (2) at 
$O(\chi^{n\ge 1})$, the {\em induced} radiative energy loss that increases 
the total radiative energy loss albeit with a reduced
efficiency due to the dead cone effect~\cite{Dokshitzer:2001zm}. The detailed
derivation of the QCD Ter-Mikayelian effect was presented in paper
II~\cite{Djordjevic:2003be}. An important conclusion from II, which 
we rely on in
the derivation below, was that the effects due to the plasmon dispersion
relation can be well approximated for high $p_T$ jets ignoring the
longitudinal modes and applying the  
asymptotic (short wavelength transverse) plasmon mass.
 That asymptotic mass, $m_g=\mu/\sqrt{2}$ is somewhat larger than the
long wavelength plasmon mass $\mu/\sqrt{3}$, where
$\mu\approx gT $ GeV is chromoelectric Debye screening mass. 

We concentrate here as in~\cite{GLV} on the case of high $p_T$ eikonal jets
produced inside a finite evolving QGP at some initial point 
$(t_0,z_0,{\bf x}_0)$. This is in contrast to the Gunion-Bertsch
problem~\cite{Bertsch-Gunion} with $t_0=-\infty$, where the jet is prepared as
a beam in the remote past. We model the interactions in the QGP as 
in~\cite{GLV,Gyulassy_Wang} via random color screened potentials 

\beqar 
V_n = V(q_n)e^{iq_n x_n} = 2\pi \delta(q^0) v(\vec{\bf q}_n) 
e^{-i \vec{\bf q}_n\cdot\vec{\bf x}_n} \; T_{a_n}(R)\otimes T_{a_n}(n) \;\; ,
\eeqar{gwmod} where $\vec{\bf x}_n$ is the location of the $n^{\rm th}$ 
scattering center and $v(\vec{\bf q}_n)\equiv 
{4\pi \alpha_s}/({\vec{\bf q}_n^{\;2}+\mu^2})$. The elastic cross section 
between the jet and target partons in the GW model is

\beqar
\frac{d\sigma_{el}(R,T)}{d^2{\bf q}}=
\frac{C_RC_2(T)}{d_A}\frac{|v({\bf q})|^2}{(2\pi)^2} \;\; . 
\eeqar{sigel}

For consistency with GLV~\cite{GLV}, we use the same notation throughout.  
Transverse 2D vectors are denoted as bold ${\bf p}$, 3D vectors as vectors 
$\vec{\bf p}=(p_z,{\bf p})$, and four vectors by $p=(p^0,\vec{\bf p})=
[p^0+p^z,p^0-p^z,{\bf p}]$. The color exchange bookkeeping with the target 
parton $n$ is handled by an appropriate $SU(N_c)$ generator~\cite{GLV}, 
$T_a(n)$, in the $d_n$ dimensional representation of the target 
(${\rm Tr} \, T_a(n)=0$ and ${\rm Tr} \,(T_a(i)T_b(j))=\delta_{ij} 
\delta_{ab} C_2(i)d_i/d_A$). We assume that all target partons are in the 
same $d_T$ dimensional representation with Casimir $C_2(T)$.) We denote the 
generators in the $d_R$ dimensional representation corresponding to the jet 
by $a\equiv t_a$ with $aa=C_R{\bf 1}$.  For heavy quarks in $SU(3)$, 
$C_R=4/3$ while $C_A=3$. The elastic cross section of target parton $i$ with 
the jet is therefore proportional to the product of Casimirs, $C_R C_2(T)$.
As in~\cite{GLV}, the analytic results derived below with the reaction 
operator approach do not depend on the actual form of $v$, but the Yukawa 
form will be used for convenience in numerical estimates. 

The sections are organized as follows: In section 2, we review the zeroth order
in opacity but in medium {\em associated} radiation for a massive quark jets.
At zeroth order only the hard initial vertex acts as a source for gluon
radiation, but in a medium the dynamical gluon mass,
$\sqrt{\omega_{pl}^2(k)-k^2}\approx m_g$ is  taken into account
~\cite{Djordjevic:2003be}.  In section 3, the first order in opacity heavy
quark energy loss is computed.  We improve Eq.~(2) in paper
I~\cite{Djordjevic:2003qk} ( see Eq.~(\ref{E1_ind}) below) by incorporating more carefully
the kinematic constraints so that the GLV massless $M\rightarrow 0$ 
limit can be recovered. The detailed
diagrammatic computations are recorded in Appendices A-G closely following the
derivation in GLV~\cite{GLV} but pointing out the different phases and residues
that arise when $M$ and $m_g$ are not zero.  Section 4, presents the
generalization of GLV to all orders in opacity for finite masses.  Our main
result (Eq.(\ref{ndifdis})), which follows from the reaction operator method
using the results from the appendices, is that the analytic structure of the
small-x induced radiation distribution in the heavy quark case is identical to
that of the massless case but with a single universal energy shift
$\Delta\omega=(m_g^2+x^2 M^2)/(2 x E)$. The energy shift is independent of 
the momentum transfers ${\bf q}_i$ and the plasmon transverse momentum 
${\bf k}$. In section 5 we give numerical estimates for the the first order 
energy loss of charm and bottom quarks comparing RHIC and LHC conditions. We 
show quantitatively how close our (partially coherent) numerical results are 
to the incoherent limit. Conclusions are presented in section 6.

\section{The one gluon associated radiation}

In order to compute the associated radiative energy loss when the hard 
process is embedded in a dielectric medium, we need to compute
the squared amplitude of Feynman diagram, $M_{rad}^{0}$, which 
represents the source $J$ that produces an off-shell jet with
momentum $p^{\prime }$ and subsequently radiates a gluon 
obeying the dispersion relation, $\omega(k)$, of the medium with momentum $k$. 
The jet emerges with momentum $p$ and mass $M$. Since our focus is on heavy 
quarks, we neglect the thermal shifts of the heavy quark mass here. As noted
before, the detailed computation of the associated energy loss was  done 
in~\cite{Djordjevic:2003be}, where it was shown that gluons in the medium can 
be approximated as massive transverse plasmons with mass 
$m_{g} \approx m_{\infty}= \mu/ \sqrt {2}$. 

If we assume that $p_{z}$ and $k_{z}$ are large enough, such 
that $M^{2}/p_{z}^{2} \ll 1$ and $m_{g}^{2}/k_{z}^{2} \ll 1$ are satisfied, 
then we can write $p$, $k$, and the transverse 
polarization $\epsilon$ in terms of light cone components:
 
\beqar
k=[xE^+,k^- 
,{\bf k}] \;\;, 
\epsilon(k)=[0,2\frac{\bf{\epsilon} \cdot
{\bf k}}{xE^+},\bf{\epsilon}]\;\;, 
\;\;  p=[(1-x)E^+,p^-,{\bf p} ] \;\;.
\eeqar{kinem}

Soft radiation is defined as $x\ll 1$ so that, for example, $p^+\gg k^+$ and 
$p^-=({\bf p}^2 + M^{2})/(1-x)E^+\ll k^-= ({\bf k}^2+m_{g}^{2})/x E^+$.
Here, $M$ is the mass of the quark, and $m_{g}$ is the mass of the glue
of energy $\omega\approx x E^+/2$. We also adopt the same shorthand notation 
as in~\cite{GLV} for energy differences:

\beqar
\omega_0=\frac{{\bf k}^{2}}{2\omega}\; ,\;
\omega_i=\frac{({\bf k}-{\bf q}_{i})^2}{2\omega}\; ,\;
\omega_{(ij)}=\frac{({\bf k}-
{\bf q}_{i}-{\bf q}_{j})^2}{2\omega} \; .
\eeqar{shorth}
In the soft eikonal kinematics that we consider
\beqar
E^+\gg k^+\gg \omega_{(i\cdots j)}+\frac{m_{g}^{2}}{2 \omega}\gg 
\frac{({\bf p}+ {\bf k})^2+M^{2}}{E^+} \;\;.
\eeqar{eorder}

The hard jet radiation amplitude to emit a transverse plasmon with
momentum, polarization, and color $(k,\epsilon,c)$ without final state
interactions is 

\beqar
M_{rad}^{0}&=&iJ(p+k)e^{i(p+k)x_0}(ig_s)(2p+k)_\mu\epsilon^\mu(k)
i\Delta_{M}(p+k)c \nonumber \\[1ex] 
&\approx& 
J(p)e^{ipx_0}(-2ig_s) \frac{\bf{\epsilon}\cdot{\bf k}}
{k^2+m_{g}^{2}+M^{2}x^{2}} \; e^{i\omega_0z_0}\; c \;\;, 
\eeqar{m0}

where $x_0(0,z_0,\bf{0})$ is the jet production point inside the plasma. We 
assume, as in~\cite{GLV}, that $J$ varies slowly with $p$, and neglect high 
$x$  spin effects. Eq.(\ref{m0}) represents the mass corrections to the 
Eg.(35) in~\cite{GLV}. In soft radiation approximation the spectrum can be 
extracted as

\beqar
|M_{rad}^{0}|^{2} \frac{d^{3} \vec{{\bf p}}}{2 E (2 \pi)^{3}}
\frac{d^{3} \vec{{\bf k}}}{2 \omega (2 \pi)^{3}} &\approx&  
d^{3}N_J d^{3} N_g^{(0)}
\; \; ,
\eeqar{imm1a} where (with $d_R=3$ dimensional representation quarks)

\beqar
d^{3} N_{J} = d_{R} |J(p)|^{2} 
\frac{d^{3}\vec{\mathbf{p}}}{( 2\pi )^{3}2p^{0}} \; \; .
\eeqar{FO2}
Eqs.~(\ref{imm1a},~\ref{FO2}) lead to the finite mass generalization of the 
small $x$ invariant DGLAP radiation spectrum 

\beqar
\omega \frac{dN_g^{(0)}}{d^3\vk} \approx x\frac{dN_g^{(0)}}{dx d^2\vk_\perp}
\approx \frac{C_R \alpha_s }{\pi^2} \frac{{\bf k}^{2}}
{({\bf k}^{2}+  m_g^2 + x^2 M^2)^{2}}\; \; .
\eeqar{imm2}
Eq.~(\ref{imm2}) 
clearly shows the depletion of radiation in the ``dead cone''~\cite{Dokshitzer:2001zm} 
at angles 
$$\theta < \theta_c=\sqrt{m_g^2 + x^2 M^2}/(xE)$$ 
generalized to take into account also the
Ter-Mikayelian dielectric dispersion in a QGP.

\section{First order radiative energy loss}

The first order in opacity energy loss can be computed from 
formula:

\beqar
d^{3} N_{g}^{(1)} d^{3} N_{J} = ( \frac{1}{d_{T}}
{\rm Tr} \left\langle |M_{1}|^{2} \right\rangle + 
\frac{2}{d_{T}}{\rm Re}{\rm Tr} \left\langle M_{0}^{\ast }M_{2}\right\rangle ) 
\frac{d^{3}\vec{\mathbf{p}}} {(2\pi )^{3}2p^{0}}
\frac{d^{3}\vec{\mathbf{k}}} {( 2\pi)^{3} 2\omega }\; \; ,
\eeqar{FO1}

where $ d^{3} N_{J}$ is given by Eq.~(\ref{FO2}) and $d_T$ is the dimension of
the target color representation ($=8$ for a pure gluon plasma). $M_{1}$ is sum 
of all diagrams with one scattering center and $M_{2}$ is sum of all diagrams 
with two scattering centers in the contact limit. We compute those diagrams 
using the same assumptions as in~\cite{GLV}, as reviewed in Appendix A.
The detailed evaluation of the amplitudes is presented in Appendices B-F.
The results are combined in appendix G, to give the small $x$ differential
energy loss

\beqar
\frac{ d E_{ind}^{(1)}}{d x} &=& \frac{C_{R} \alpha_{S}}{\pi} 
\frac{L}{\lambda} E \int \frac {d \mathbf{k}^{2}}{\mathbf{k}^{2} + 
m_{g}^{2}+M^{2}x^{2}} \int \frac{d^{2} \mathbf{q}_{1}} {\pi} 
\frac{\mu^{2}}{(\mathbf{q}_{1}^2 +\mu^{2})^{2}} \times \nonumber \\[1ex]
&\times& 2 \; \; \frac{ \mathbf{k} \cdot \mathbf{q}_{1} 
( \mathbf{k}-\mathbf{q}_{1})^{2} + (m_{g}^{2}+M^{2}x^{2})\; \mathbf{q}_{1} 
\cdot (\mathbf{q}_{1}-\mathbf{k})}{( \frac{4 E x}{L} )^{2} 
+ (( \mathbf{k}-\mathbf{q}_{1})^{2} + M^{2}x^{2} + m_{g}^{2})^{2}} \; \; , 
\eeqar{dE/dx1}

where $|\mathbf{q}_{1}|$ is the magnitude of the transverse momentum transfer 
between a target parton and a jet, $|\mathbf{k}|$ is the magnitude of the 
transverse momentum of the radiated gluon, and $\lambda_g$ is the mean free 
path of the gluon. The simple analytic form of the destructive interference
factor involving $(4 E x/L)^2$ arises for an assumed exponential 
distribution $\exp(-\Delta z/L)/L$ of the distance between the jet
production and target rescattering center. 
Note that in the massless limit, $M=m_{g}=0$, Eq.~(\ref{dE/dx1}) reduces to 
Eq. (125) in~\cite{GLV}.
\medskip 
 
Numerical evaluation of Eq.\ref{dE/dx1} shows that
to a good approximation we can ignore the finite kinematic bounds on
$|\mathbf{q}_{1}|<\sqrt{6ET}$ for $E>10$ GeV charm jets. 
With this simplification,
we can substitute $\mathbf{q} \rightarrow \mathbf{q}_{1}-\mathbf{k}$
and integrate over the azimuthal angle to arrive at

\beqar
\Delta E^{(1)}_{ind} &=& \frac{C_{F}\alpha_{S}}{\pi} \frac{L}{\lambda_g}
\int_0^1 dxE \int_0^\infty 
\frac{ 2 \mathbf{q}^2 \mu^2 d\mathbf{q}^2}{( \frac{4 E x}{L} )^{2} 
+ (\mathbf{q}^{2} + M^{2}x^{2} + m_{g}^{2})^{2}} \nonumber \\
&\; & \int \frac{ d\mathbf{k}^2 \; \theta (x E-|\mathbf{k}|)}
{(( |\mathbf{k}|-|\mathbf{q}|)^{2} + \mu^{2})^{3/2} 
(( |\mathbf{k}|+|\mathbf{q}|)^{2} + \mu^{2})^{3/2}} 
\left\{ \mu^2+ (\mathbf{k}^{2}-\mathbf{q}^{2}) 
\frac{\mathbf{k}^{2} - M^{2}x^{2} - m_{g}^{2}}
{\mathbf{k}^{2} + M^{2}x^{2} + m_{g}^{2}} \right\}, 
\eeqar{E1_ind}

If we further neglect the finite kinematic boundaries on the gluon transverse
momentum, 
$|\mathbf{k}|< xE$, then in the massless $M=m_{g}=0$ limit, 
we recover the approximate asymptotic 
Eq.~(127) of \cite{GLV}. The $\mathbf{k}^2$ integral in Eq.~(\ref{E1_ind}) 
can be performed analytically if $\alpha_{S}$ does not run, but the cumbersome
result is not instructive. 

We note that Eq.~(\ref{E1_ind}) differs from the Eg.~(2) 
in~\cite{Djordjevic:2003be}. The $\Delta E^{(1)}_{ind}$ 
in~\cite{Djordjevic:2003be} was obtained by assuming that energy 
of the glue is much larger than momentum transfer $\mathbf{q}_{1}$, i.e 
$xE \pm |\mathbf{q}_{1}| \approx xE$. Eq.~(\ref{E1_ind}) avoids this approximation
by incorporating exact kinematics at the price  of a more complicated expression.
Numerically, this improvement however, does not 
change significantly the estimated charm energy loss presented
in ~\cite{Djordjevic:2003be} as shown in section 5.
However, theoretically, Eq.~(\ref{E1_ind}) is prefered since the $M\rightarrow 0$ 
limit is only correctly recovered when exact kinematics are enforced.

The main qualitative effect of increasing $M$ is to reduce the 
relevance of the inverse 
formation time factor $(4 E x/L)^2$ in the denominator of 
the integrand. Formally, by setting this factor to zero recovers the 
incoherent limit of induced radiation.
This corresponds to the QCD analog of the QED Bethe-Heitler limit
with $\Delta E_{GB} \propto \alpha_s E L /\lambda_g$ 
modulo a logarithmic factor. In QCD this includes the isolated scattering
Gunion-Bertsch (GB) radiation as well as elastic scattering of the
associated radiation from zeroth order in opacity as we discuss in the following section.

The incoherent, short formation time limit, generalized here to include the 
plasmon asymptotic mass,  is  given by
\beqar
\frac{\Delta E_{in}}{E} &=& \frac{C_{F}\alpha_{S}}{\pi} \frac{L}{\lambda_g}
\int_0^1 dx \int 
\frac{ 2 \mathbf{q}^2 \mu^2 d\mathbf{q}^2}
{(\mathbf{q}^{2} + M^{2}x^{2} + m_{g}^{2})^{2}} \nonumber \\
&\; & \int \frac{ d\mathbf{k}^2 \; \theta (x E-|\mathbf{k}|)}
{(( |\mathbf{k}|-|\mathbf{q}|)^{2} + \mu^{2})^{3/2} 
(( |\mathbf{k}|+|\mathbf{q}|)^{2} + \mu^{2})^{3/2}} 
\left\{ \mu^2+ (\mathbf{k}^{2}-\mathbf{q}^{2}) 
\frac{\mathbf{k}^{2} - M^{2}x^{2} - m_{g}^{2}}
{\mathbf{k}^{2} + M^{2}x^{2} + m_{g}^{2}} \right\}, 
\eeqar{E1_in}

Since we found in Ref.~\cite{Djordjevic:2003qk}, and show in more detail in 
section 5 that the heavy quark energy loss is close to the incoherent 
limit, it is also useful to define the effective gluon radiation length, 
$L_{rad}$, via

\beqar
\frac{d\Delta E}{d L} \equiv \frac{E}{L_{rad}(M,m_g,E,L) }
\; \; . 
\eeqar{Lrad}

This length is of relevance to  answer the question whether the jets observed 
in a given kinematic window are dominated by surface or volume emission.

\section{Higher orders in opacity energy loss}

\subsection{Heavy quark generalization of GLV to all orders}
From the results of the calculations reported in the appendices, we find 
that the finite masses modify the effective radiation amplitudes and phase 
factors in a remarkably simple and universal way. The phase factors in the 
massless case studied in GLV~\cite{GLV} modulate the amplitudes by factors 
such as $Exp(i \sum_m \omega_{(m\cdots n)}\Delta z_m)$. The energy 
differences $\omega_{(m\cdots n)}$ from Eq.(\ref{shorth}) are inverse 
formation times and $\Delta z_m$ are distances between scattering centers.
With finite $M$ and $m_g$ these energy differences are found from 
Eqs.~(\ref{101b}, \ref{100final}, \ref{110final}, \ref{203d}, \ref{200fin},
\ref{220fin}, \ref{201recov}) to be simply shifted by a $q_i$ independent 
term, 

\beqar
\omega_{(m,\cdots , n)}=
\frac{({\bf k}-{\bf q}_m-\cdots-{\bf q}_n)^2}{2 x E} 
\rightarrow \Omega_{(m,\cdots , n)}\equiv \omega_{(m,\cdots , n)}+ 
\frac{m_{g}^{2} +M^{2}x^{2}}{2 x E}
\;\;  
\eeqar{omegm}

In addition, the kinematic current amplitudes appearing in those equations are 
simply modified versions of the Hard, Gluon-Bertsch and Cascade terms in GLV, 
which for finite masses are now 

\beqar
{\bf \tilde{H}}&=& \frac{{\bf k}}{{\bf k}^2 + m_{g}^{2} +M^{2}x^{2}}\; , 
\qquad \qquad {\bf \tilde{C}}_{(i_1i_2 \cdots i_m)}=
\frac{({\bf k} - {\bf q}_{i_1} - {\bf q}_{i_2}- \cdots -{\bf q}_{i_m} )}
{({\bf k} - {\bf q}_{i_1} - {\bf q}_{i_2}- \cdots -{\bf q}_{i_m})^2
+ m_{g}^{2} +M^{2}x^{2}} \;, 
\nonumber \\[1.ex]
{\bf \tilde{B}}_i &= &{\bf \tilde{H}} - {\bf \tilde{C}}_i \; , \qquad \qquad 
\qquad \qquad \;
{\bf \tilde{B}}_{(i_1 i_2 \cdots i_m )(j_1j_2 \cdots i_n)} = 
{\bf \tilde{C}}_{(i_1 i_2 \cdots j_m)} - {\bf \tilde{C}}_{(j_1 j_2 \cdots j_n)}\; \; .
\eeqar{hbgcdef}

In summary, the computations in Appendices B-F show that the diagrams for the 
finite masses case can be obtained from corresponding massless equations 
in~\cite{GLV}, by simply replacing the terms from Eqs.~(58, 106) in~\cite{GLV}
by the modified terms defined in Eqs.~(\ref{hbgcdef},~\ref{omegm}). 

Therefore, the recursive GLV reaction operator formalism 
carries over to the massive case with the simple replacements above.
The complete arbitrary order in opacity induced radiation distribution
can be obtained by generalizing  Eq.(113) of GLV as follows:

\beqar
x\frac{dN^{(n)}}{dx\, d^2 {\bf k}} &=&
\frac{C_R \alpha_s}{\pi^2} \frac{1}{n!} 
\int \prod_{i=1}^n \left(d^2{\bf q}_{i}\,
\frac{L}{\lambda_g(i)} 
\left[\bar{v}_i^2({\bf q}_{i}) - \delta^2({\bf q}_{i}) \right]\right) 
\, \times \nonumber \\[1.ex]
&\;& \times \left( -2\,{\bf \tilde{C}}_{(1, \cdots ,n)} \cdot 
\sum_{m=1}^n {\bf \tilde{B}}_{(m+1, \cdots ,n)(m, \cdots, n)} 
\left[ \cos \left (\, \sum_{k=2}^m \Omega_{(k,\cdots,n)} \Delta z_k \right)
-   \cos \left (\, \sum_{k=1}^m \Omega_{(k,\cdots,n)} \Delta z_k \right)
\right]\; \right) \;\;, \nonumber   \\
\nonumber \\
\eeqar{ndifdis} 

where $\sum_2^1 \equiv 0$ is understood and $|\bar{v}_i({\bf q}_{i}) |^2$ is 
defined as the normalized distribution of momentum transfers from 
$i^{{\rm th}}$ scattering center~\cite{GLV}. For the Yukawa screened
interactions (Eq.~(\ref{sigel})), the differential 
gluon cross section in the local density approximation is 
\beqar
\sigma(z_i,{\bf q}_i)\equiv \sigma_{el}(z_i)|\bar{v}_i({\bf q}_{i})|^2=
\frac{d^2 \sigma_{el}(z_i)}{d^2{\bf q}_{i}}=\frac{\sigma_{el}(z_i)}{\pi}
\frac{\mu(z_i)^2}{({\bf q}^{2}+\mu(z_i)^2)^2}
\;\; ,
\eeqar{vbar}
where $\mu(z_i)$ is the local Debye mass that may vary if the system expands.

Note that, as in~\cite{GLV} the Eq.~(\ref{ndifdis}) is not restricted to 
uncorrelated geometries. Also it allows the inclusion of finite kinematic 
boundaries on the ${\bf q}_i$ as well as different functional forms of the 
gluon elastic cross sections  along the eikonal path. The $n^{{\rm th}}$ order
in opacity energy loss spectrum can be obtained from Eq.~(\ref{ndifdis}) via

\beqar
\frac{dE^{(n)}}{dx} = \int \, d^2 {\bf k} \frac{dN^{(n)}}{dx\, d^2 {\bf k}} 
\;xE
\eeqar{En}

From the Eqs.~(\ref{ndifdis},~\ref{En}) we can obtain the first order 
in opacity energy loss by setting $n=1$. The result obtained after averaging
over $\exp(-\Delta z_1/L)/L$ is the same as Eq.~(\ref{dE/dx}) in Appendix G, 
which leads to Eq.~(\ref{E1_ind}) computed in previous section.

In order to make the averaging over the target coordinates more explicit, 
assume an uncorrelated geometry and let $\rho(z)\equiv\rho(z,\tau=z)$ denote 
the target density along the path of the jet.  For Bjorken expansion $\rho(z)=
\theta(R-z) \rho_0 \tau_0/z$ for example. In the local density approximation, 
the screening mass, $\mu(z)$, may also depend on the proper time $(\tau=z)$. 
Then $\sigma_{el}(z)$ and $|\bar{v}(z,{\bf q})|^2$ may vary along the jet 
path as well. The average over over the target centers can be made explicit 
by replacing the opacity factor in (\ref{ndifdis}) by

\beqar
\frac{1}{n!}
\int \prod_{i=1}^n \left(d^2{\bf q}_{i}\,
\frac{L}{\lambda_g(i)}\left[\bar{v}_i^2({\bf q}_{i}) - \delta^2({\bf q}_{i}) 
\right]\right) &\rightarrow & \int_0^\infty dz_1 \rho(z_1) \cdots
\int_{z_{n-1}}^\infty dz_n \rho(z_n) \times
\nonumber \\
&\;& \hspace{0.3in} \int\prod_{i=1}^n \left( d^2{\bf q}_{i}\,
\left[ \sigma(z,{\bf q}_{i})- \sigma_{el}(z)\delta^2({\bf q}_{i}) \right]
\right)
\eeqar{zint}

\subsection{The Incoherent, Short Formation Time Limit}

The incoherent limit of Eq.(\ref{ndifdis}) is obtained formally by taking the 
large $\Omega_{(k,\cdots,n)} \Delta z_k \gg 1$ limit in which all 
$\cos(\cdot)$ factors average to zero except for one term for $m=1$ in which 
the $\cos(0)=1$. Hence the incoherent limit of the $n^{th}$ order in opacity 
induced radiation is

\beqar
x\frac{dN^{(n)}_{in}}{dx\, d^2 {\bf k}} &=& 
\frac{C_R \alpha_s}{\pi^2} \frac{1}{n!}  \int \prod_{i=1}^n 
\left(d^2{\bf q}_{i}\, \left(\frac{L}{\lambda_g(i)}\right)
\left[\bar{v}_i^2({\bf q}_{i}) - \delta^2({\bf q}_{i}) \right]\right) 
\left( -2\,{\bf \tilde{C}}_{(1, \cdots ,n)} \cdot 
{\bf \tilde{B}}_{(2, \cdots ,n)(1, \cdots, n)} \right) \;\;, 
\eeqar{GBndis} 

For the $n=1$ contribution

\beqar
x\frac{dN^{(1)}_{in}}{dx\, d^2 {\bf k}} &=& \frac{C_R \alpha_s}{\pi^2} 
\left( \frac{L}{\lambda_g} \right) \int d^2{\bf q}_{1}\, 
\left[\bar{v}^2({\bf q}_{1}) - \delta^2({\bf q}_{1}) \right]
\left( -2\,{\bf \tilde{C}}_{(1)} \cdot {\bf \tilde{B}}_{(0)(1)} \right)  
\nonumber   \\
&=& \frac{C_R \alpha_s}{\pi^2} \left( \frac{L}{\lambda_g} \right)
\int d^2{\bf q}_{1}\, \bar{v}^2({\bf q}_{1}) 
\left( {\bf \tilde{B}}_{(0)(1)}^2+ {\bf \tilde{C}}_{(1)}^2-{\bf \tilde{H}}^2
\right) \;\;, 
\eeqar{GB1dis} 

where ${\bf \tilde{B}}_{(0)(1)}={\bf \tilde{H}}-{\bf \tilde{C}}_{(1)}$ is the 
finite mass generalization of the incoherent GB radiation amplitude including 
the asymptotic plasmon dispersion. Note that (see Wiedemann~\cite{ELOSS}) 
the $-{\bf \tilde{H}}^2$ term corresponds to the first order unitarity
correction of the full Glauber {\em associated} zeroth order in opacity 
contribution $e^{-L/\lambda_g} {dN^{(0)}}/{dx\, d^2 {\bf k}}$. In addition 
elastic rescattering of the radiated gluon leads to the cascade contribution 
${\bf \tilde{C}}_{(1)}^2$ which further broadens the (already very broad) 
transverse momentum distribution of the associated radiation. The incoherent 
induced radiation from the rescattering of the quark jet in the medium is 
the Gunion-Bertsch term ${\bf \tilde{B}}_{(0)(1)}^2$. There are $L/\lambda_g$ 
such induced contributions in the completely incoherent limit.

In general we expect that with the inclusion of masses in Eq.(\ref{omegm}), 
the increase of the  $\Omega_{(m, \cdots , n)} \Delta z_m$ arguments of the 
interference cosines in Eq.(\ref{ndifdis}) will drive the radiation 
distribution closer to the incoherent limit. In the next section we will 
evaluate numerically how close is charm and bottom quark induced radiation 
to their respective incoherent  limits. 

Similar to the massless case studied in GLV~\cite{GLV}, we expect that in the 
case of finite mass, higher orders in opacity contributions to the net 
induced  energy loss will converge rapidly for moderate $L/\lambda_g$ of 
practical interest. The dominance of the first order contribution to the 
transverse momentum integrated energy loss is more readily seen in the 
incoherent limit above. Change variables ${\bf k} \rightarrow {\bf k}'\equiv 
{\bf k} -{\bf q}_2 -\cdots -{\bf q}_n$ and integrate over ${\bf k}'$.
In the $\omega \gg \mu$ limit, we can approximately ignore the change in the 
$|{\bf k}'|<\omega$ kinematic limit. In that case

\beqar
\int d^2{\bf k} \; {\bf \tilde{C}}_{(1, \cdots ,n)} \cdot 
{\bf \tilde{B}}_{(2, \cdots ,n)(1, \cdots, n)} \approx \int d^2{\bf k}' 
\; {\bf \tilde{C}}_{(1)} \cdot {\bf \tilde{B}}_{(0)(1)} \; \; .
\eeqar{kintBG}

However, the integrated energy loss at order $n\ge 2$ has additional
integrations 

$$\int \prod_{i=2}^n d^2{\bf q}_{i}\left[\bar{v}_i^2({\bf q}_{i}) - 
\delta^2({\bf q}_{i}) \right]=0$$

that tend to kill the whole contribution modulo tiny edge effects due to 
Eq.(\ref{kintBG}) not being exact. Therefore, especially as we approach the 
incoherent limit the main order contribution to the medium induced energy 
loss is dominated by the first order in opacity. 

In order to see how this works in more detail consider the case of 
uncorrelated geometries where we can use Eq.(\ref{zint}) and Fourier 
techniques to sum the whole opacity series in this incoherent limit. Note 
first that the term proportional to $\delta^2({\bf q}_{1})$ vanishes due to 
${\bf \tilde{B}}_{(2, \cdots ,n)(1, \cdots, n)}=0$ when ${\bf q}_{1}=0$.
It is convenient therefore to define the accumulated 
${\bf Q}=\sum_2^n{\bf q}_{i}$ momentum transfer distribution by inserting a 
factor 
$$ 1=\int d^2{\bf Q} \delta^2({\bf Q}-\sum_{i=2}^n{\bf q}_{i})=
\int \frac{d^2{\bf Q}}{(2\pi)^2}\; \int d^2{\bf b}\; 
e^{i {\bf Q}\cdot{\bf b}} \; \prod_{i=2}^n e^{-i {\bf q}_i\cdot{\bf b}} $$

Define an effective dipole cross section
$$\sigma_d(z,{\bf b})\equiv
-\int d^2{\bf q} \; e^{-i{\bf q}\cdot{\bf b}} 
\left(\sigma(z_i,{\bf q}_{i}) - \sigma_{el}(z)\delta^2({\bf q}_{i}) \right)$$
and a corresponding dipole opacity $$\chi_d(z,{\bf b})\equiv \rho(z_i)
\sigma_d(z,{\bf b}). $$

In terms of these quantities,
\beqar
x\frac{dN^{(n)}_{in}}{dx\, d^2 {\bf k}} &=& \frac{C_R \alpha_s}{\pi^2} 
\int_0^\infty dz_1 d^2{\bf q}_{1} \rho(z_1)\sigma(z_1,{\bf q}_{1}) 
\nonumber \\ 
&\;& \int \frac{d^2{\bf Q}}{(2\pi)^2} \; \int d^2{\bf b}\;
e^{i {\bf Q}\cdot{\bf b}} \;  \int_{z_1}^\infty dz_2 \; 
(-1)\chi_d(z_2,b)\int_{z_3}^\infty \cdots \int_{z_{n-1}}^\infty 
dz_n \; (-1)\chi_d(z_n,b)\, \times \nonumber \\[1.ex]
&\;& \hspace{2in} \times  2 \,{\bf \tilde{C}}_{{\bf k - Q - q}_1} \cdot 
\left( {\bf \tilde{C}}_{{\bf k - Q - q}_1} -  
{\bf \tilde{C}}_{\bf k-Q} \right) 
\nonumber \\
&=& \frac{C_R \alpha_s}{\pi^2} \int_0^\infty dz_1 d^2{\bf q}_{1} \rho(z_1) 
\sigma(z_1,{\bf q}_{1}) \nonumber \\ 
&\;& \int \frac{d^2{\bf Q}}{(2\pi)^2}
\; \int d^2{\bf b}\;
e^{i {\bf Q}\cdot{\bf b}} \; \frac{(-1)^{n-1}}{(n-1)!}\left(
 \int_{z_1}^\infty dz_2 \chi_d(z_2,b) \right)^{n-1}
\, \times \nonumber \\[1.ex]
&\;& \hspace{2in} \times  2 \,{\bf \tilde{C}}_{{\bf k - Q - q}_1} \cdot 
\left( {\bf \tilde{C}}_{{\bf k - Q - q}_1} -  
{\bf \tilde{C}}_{\bf k-Q} \right) 
\eeqar{GBndis2} 

We can now sum all orders from $n=1,\infty$ into a closed form
\beqar
x\frac{dN_{in}}{dx\, d^2 {\bf k}} &=&
\frac{C_R \alpha_s}{\pi^2} 
\int_0^\infty dz_1 d^2{\bf q}_{1} \rho(z_1)\sigma(z_1,{\bf q}_{1})
\nonumber \\ 
&\;& \int \frac{d^2{\bf Q}}{(2\pi)^2}
\; \left[\int d^2{\bf b}\;
e^{i {\bf Q}\cdot{\bf b}} \; e^{-\int_{z_1}^\infty dz' \rho(z')\sigma_d(z',b)} \right]
\, \times \nonumber \\[1.ex]
&\;& \hspace{2in} \times  2 \,{\bf \tilde{C}}_{{\bf k - Q - q}_1} \cdot 
\left( {\bf \tilde{C}}_{{\bf k - Q - q}_1} -  
{\bf \tilde{C}}_{\bf k-Q} \right) 
\eeqar{GBndis3} 

We can now clearly see how (\ref{GBndis3}) reduces to the first order result
when integrated over ${\bf k}$. If we change variables to 
${\bf k}'={\bf k}-{\bf Q}$ and assume that $k'<\omega$ remains approximately 
valid, then the ${\bf Q}$ integration reduces to a $\delta^2({\bf b})$ which 
in turn converts $\sigma_d(z,b)\rightarrow \sigma_d(z,0)=0$. The whole second 
line reduces then to a unit factor so that

\beqar
x\frac{dN_{in}}{dx} &\approx&
\frac{C_R \alpha_s}{\pi^2} 
\int_0^\infty dz d^2{\bf q} \rho(z)\sigma(z,{\bf q})\int d^2{\bf k}
\; 2 \,{\bf \tilde{C}}_{{\bf k- q}_{1}} \cdot 
\left( {\bf \tilde{C}}_{{\bf k-q}_{1}} - {\bf \tilde{H}} \right) 
\eeqar{GBndis4} 
However, in the same approximation that led to (\ref{GBndis4}) from
 (\ref{GBndis3}), we can again shift variables to ${\bf k}'={\bf k}-{\bf q}_1$
and again neglect the change in the $|k'|<\omega$ bound to cancel the ${\bf \tilde{C}}^2-{\bf \tilde{H}}^2$
contribution. This collapses the full result to the simple incoherent 
Gunion-Bertsch limit
\beqar
x\frac{dN_{in}}{dx} &\approx& x\frac{dN_{GB}}{dx}=
\frac{C_R \alpha_s}{\pi^2} 
\int_0^\infty dz d^2{\bf q} \rho(z)\sigma(z,{\bf q})\int d^2{\bf k}
\; {\bf \tilde{B}}_{(0)(1)}^2 
\eeqar{GBndis5}
In order to assess the accuracy of the approximations involved in
the approximate handling of the kinematic $k$ bounds, we have to evaluate
numerically the difference between $dN_{in}$ and $dN_{BG}$.
 
\section{Numerical estimates}

\subsection{Heavy quark energy loss at RHIC}
The numerical results for the first order induced radiative energy
loss are shown on Fig.1 for charm and bottom quarks. From the analysis
of light quark quenching in $Au+Au$ at $130$ GeV~\cite{Levai_opacity} 
effective static plasma opacity $L/\lambda$ is in the range $3-4$. On Fig.1 
we fix opacity to $L/\lambda=4$, and we look at the $1^{st}$ order in opacity 
fractional energy loss as a function of initial energy of the quark. We 
assume here that $\alpha_s=0.3,\; \mu=0.5\;{\rm GeV},$ and $\lambda=1$ fm 
for the plasma parameters. Since finite parton masses shield collinear 
${\bf k} \rightarrow 0$ singularity~\cite{Djordjevic:2003be}, our 
numerical computations are performed with zero momentum cutoff. We see that 
for heavy quarks, in the energy range $E \sim 5-15$ GeV, the Ter-Mikayelian 
effect reduces the induced energy loss in this extension of the GLV approach 
somewhat more than in the BDMS approximation~\cite{Dokshitzer:2001zm}. 
However, on an absolute scale, this only corresponds to a change of
$\delta(\Delta E^{(1)}/E) < 0.05$, which is negligible. Note that
with both dead cone and plasmon mass reduction, there remains a
sizeable induced energy loss fraction $\Delta E^{(1)}/E \approx 0.15$
for charm quarks while only about half that is predicted for bottom.

\begin{center}
\vspace*{6.8cm} \includegraphics{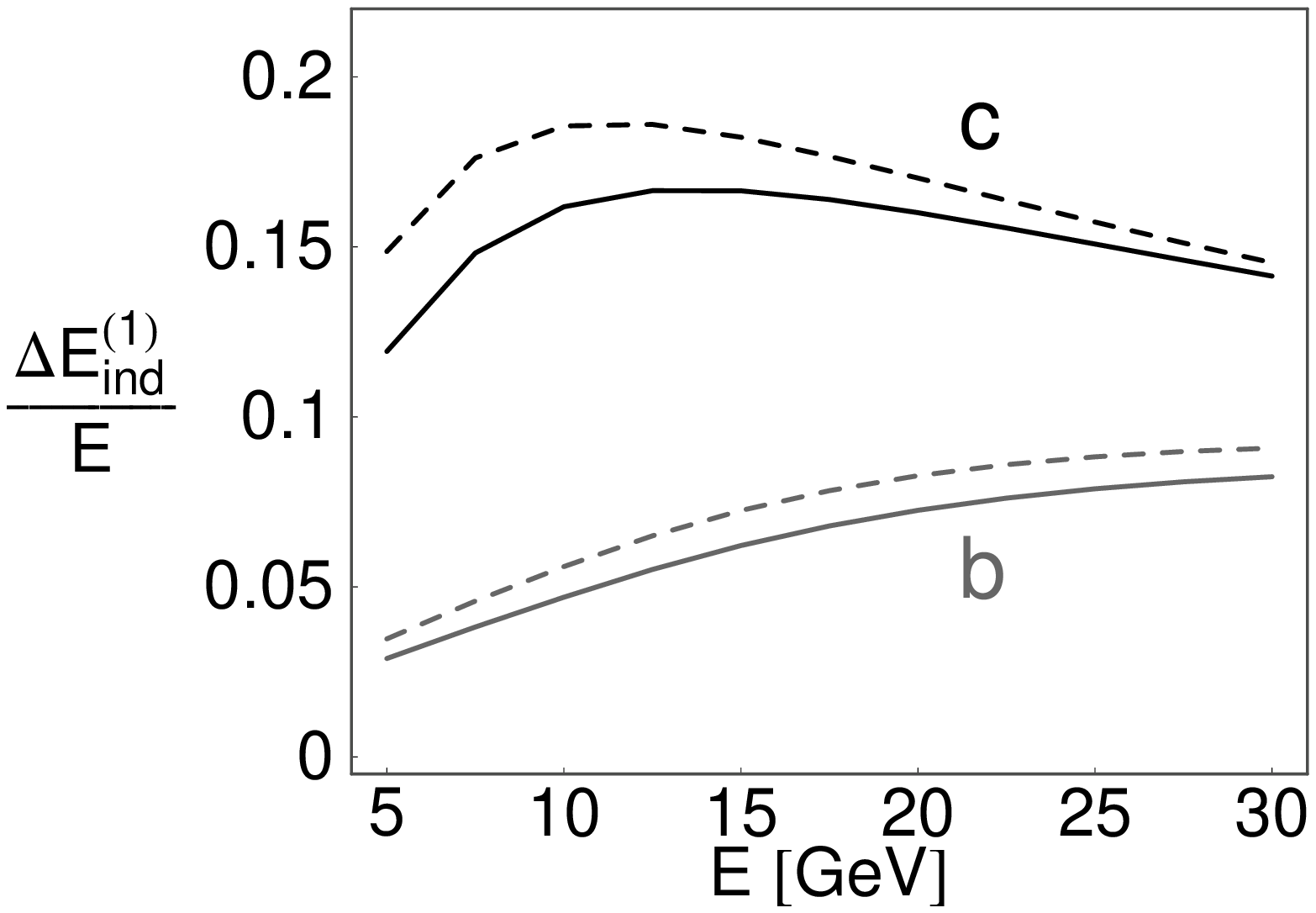}
\begin{minipage}[t]{15.0cm}
{\small {FIG~1.} The $1^{st}$ order in opacity fractional energy loss for 
heavy quarks with (solid curves) and without (dashed curves) Ter-Mikayelian 
effect approximated by a constant $m_g=\mu/\sqrt{2}$ for transverse modes 
only. Upper curves correspond to charm, and lower to bottom quarks as a 
function of their energy in a plasma characterized by $\alpha_s=0.3, \; 
\mu=0.5\;{\rm GeV},$ and $L=4\lambda=4$ fm. }
\end{minipage}
\end{center}
\vskip 4truemm 

However, we want to emphasize that, though Ter-Mikayelian effect is not 
important for the medium induced energy loss of the massive quarks, it is 
very important for massless quark case. Note that excluding Ter-Mikayelian 
effect means setting both plasmon mass and momentum cutoff to zero. Since in 
the massless parton case,  ${\bf k} \rightarrow 0$ singularities are not 
naturally regulated, excluding the Ter-Mikayelian effect would lead to 
infinite result of medium induced energy loss. In~\cite{GLV} these 
divergences are prevented by introducing the finite $|{\bf k}|>\mu$ cutoff. 
It is easy to see from numerical computations that medium induced energy 
loss for light quarks with finite plasmon mass $m_g=\mu/\sqrt{2}$ and zero 
momentum cutoff, is similar to the energy loss obtained with zero plasmon 
mass and finite momentum cutoff $|{\bf k}|>\mu$. Setting the finite momentum 
cutoff~\cite{GLV} in the case of light partons thus numerically produces 
similar results as introducing the finite plasmon mass, i.e. including the 
Ter-Mikayelian effect. Therefore, though the Ter-Mikayelian effect was not 
considered in~\cite{GLV}, the same effect was obtained by including the 
finite momentum cutoff. However, we would like to note that, though both 
approaches give similar numerical results, the advantage of the 
Ter-Mikayelian effect is that it provides the natural regulation of 
${\bf k} \rightarrow 0$ divergences.

Fig. 2 shows the first order induced radiative energy loss for charm and 
bottom quarks as a function of opacity. We see that the induced contribution 
increases with $L$. Note that for heavy (charm and bottom) quarks the 
thickness dependence is closer to the linear Bethe-Heitler like form, $L^1$, 
than the asymptotic energy quadratic form~\cite{ELOSS,GLV}.

\begin{center}
\vspace*{6.5cm} \includegraphics{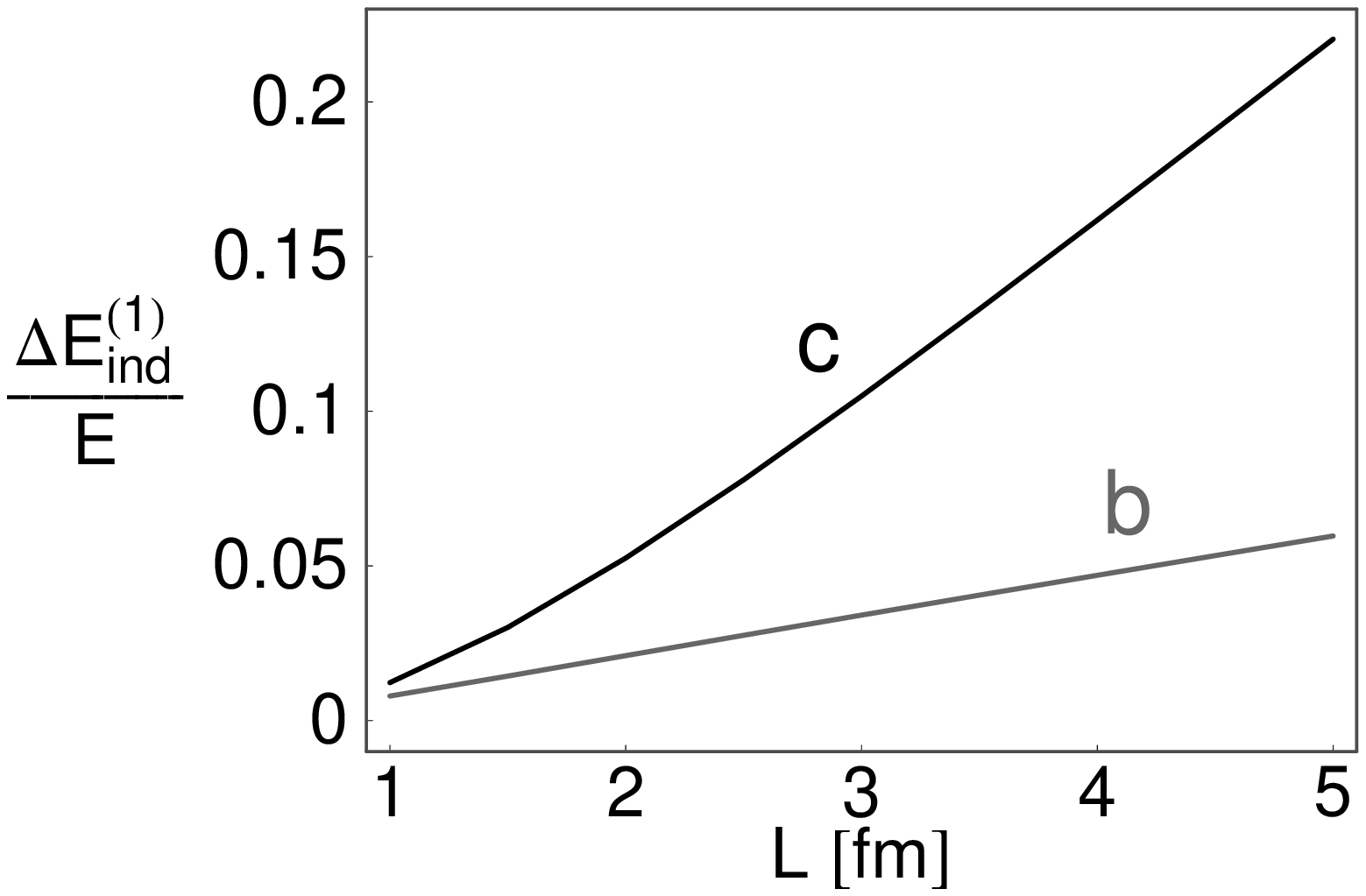}
\begin{minipage}[t]{15.0cm}
{\small {FIG~2.} The $1^{st}$ order in opacity fractional energy loss for a 
10 GeV charm and bottom quark is plotted versus the effective static 
thickness $L$ of a plasma characterized by $\mu=0.5$ GeV and $\lambda=1$ fm. 
Upper curve correspond to charm, and lower to bottom quark.}
\end{minipage}
\end{center}
\vskip 4truemm 

In the case of finite kinematic bounds on $|\mathbf{q}_{1}|<\sqrt{6ET}$, the 
angular integration in Eq.~(\ref{dE/dx1}) can still be performed, but the 
result is cumbersome. However, including finite kinematic bound on 
$|\mathbf{q}_{1}|$ is found to be unimportant, because it only reduces the induced 
energy loss fraction $\Delta E^{(1)}/E$ in Fig. 1 by less than $10 \%$.
Therefore, we can safely neglect the finite 
kinematic boundaries on the momentum transfers.

\bigskip

\begin{center}
\vspace*{7.0cm} \includegraphics{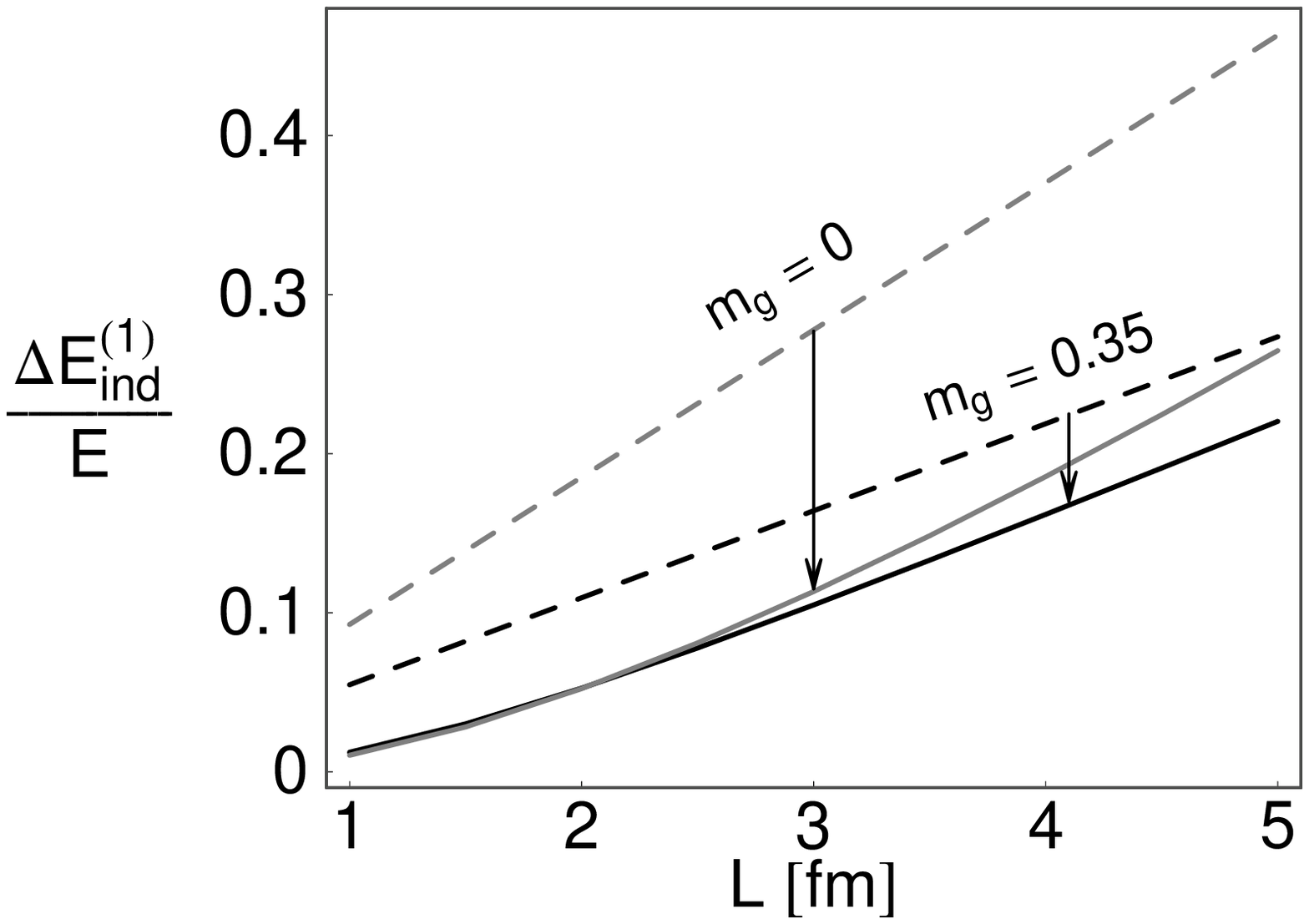}
\begin{minipage}[t]{15.0cm}
{\small {FIG~3.} Comparison of the incoherent limit 
(dashed curves) to the $1^{st}$ order (solid curves) in opacity fractional 
energy loss for $E=10$ GeV charm quarks via Eq.~(\protect{\ref{E1_ind}}) 
under RHIC conditions ($\alpha_s=0.3, \; \mu=0.5\;{\rm GeV}$). The higher 
dashed and solid curves correspond to $m_g=0$, while the lower to 
$m_g=\mu/\sqrt{2}$. Arrows point between corresponding incoherent limit 
and full $1^{st}$ order results.}
\end{minipage}
\end{center}
\vskip 4truemm 

In Fig.~3 the first order energy loss for charm quarks is compared to the 
incoherent limit obtained from Eq.(\ref{E1_ind}) by 
setting $4xE/L=0$. This analogue of the QED Bethe-Heitler limit is the one  
where no destructive LPM interference effects occur in QCD. The 
characteristic of the incoherent  limit is the complete linear dependence of 
$\Delta E$ on $L$. The slope depends on the assumed $\mu$ as well as the 
assumed dynamical gluon mass. Two cases, $m_g=0, \mu/\sqrt{2}$ are shown by 
the two dashed lines. Comparing to the solid lines that include destructive 
interference effects due to finite formation times, 
we see that the realistic case with finite $m_g=0.35 $ GeV  is 
remarkably close to the incoherent limit for $E=10$ GeV quarks.

We have also checked numerically the difference between the full first order
incoherent limit and the isolated Gunion-Bertsch contribution. These should 
be approximately the same if the elastic rescattering and the hard unitarity 
correction terms cancel as in Eq.(\ref{GBndis5}). We find that the GB energy 
loss for E=10 GeV charm quarks is $\Delta E_{GB}/E\approx 0.35$ is 
significantly larger than the incoherent energy loss 
$\Delta E_{in}/E\approx 0.22$. However by $E= 30$ GeV, the GB limit exceeds 
the incoherent limit by only $15\%$.

Finally, by using the results for associated energy loss 
from~\cite{Djordjevic:2003be}, we can now compare the net 
($\Delta E^{(1)}_{ind}+\Delta E^{(0)}_{med}$) energy loss in the medium, with 
the one in the vacuum defined by $\mu_{vac}=0$ GeV, as shown on Fig. 4. As in 
paper I, we see that even in the absence of a medium ($L=0)$, a charm quark 
with energy $\sim 10$ GeV suffers an average energy loss, 
$\Delta E^{(0)}_{vac}/E\approx 1/3$, due to the sudden change of the color 
current when it is formed in the vacuum. The dielectric plasmon effect 
reduces this to about $\Delta E^{(0)}_{med}/E\approx 1/4$. This contribution 
is independent of the thickness of the plasma as long as $L$ is not too 
small. For very small $L< 1/m_g$, the plasmon dispersion is smeared out due 
to the uncertainty principle, and $\Delta E^{(0)}_{med}$ must approach
$\Delta E^{(0)}_{vac}$ from below.

\begin{center}
\vspace*{6.9cm} \includegraphics{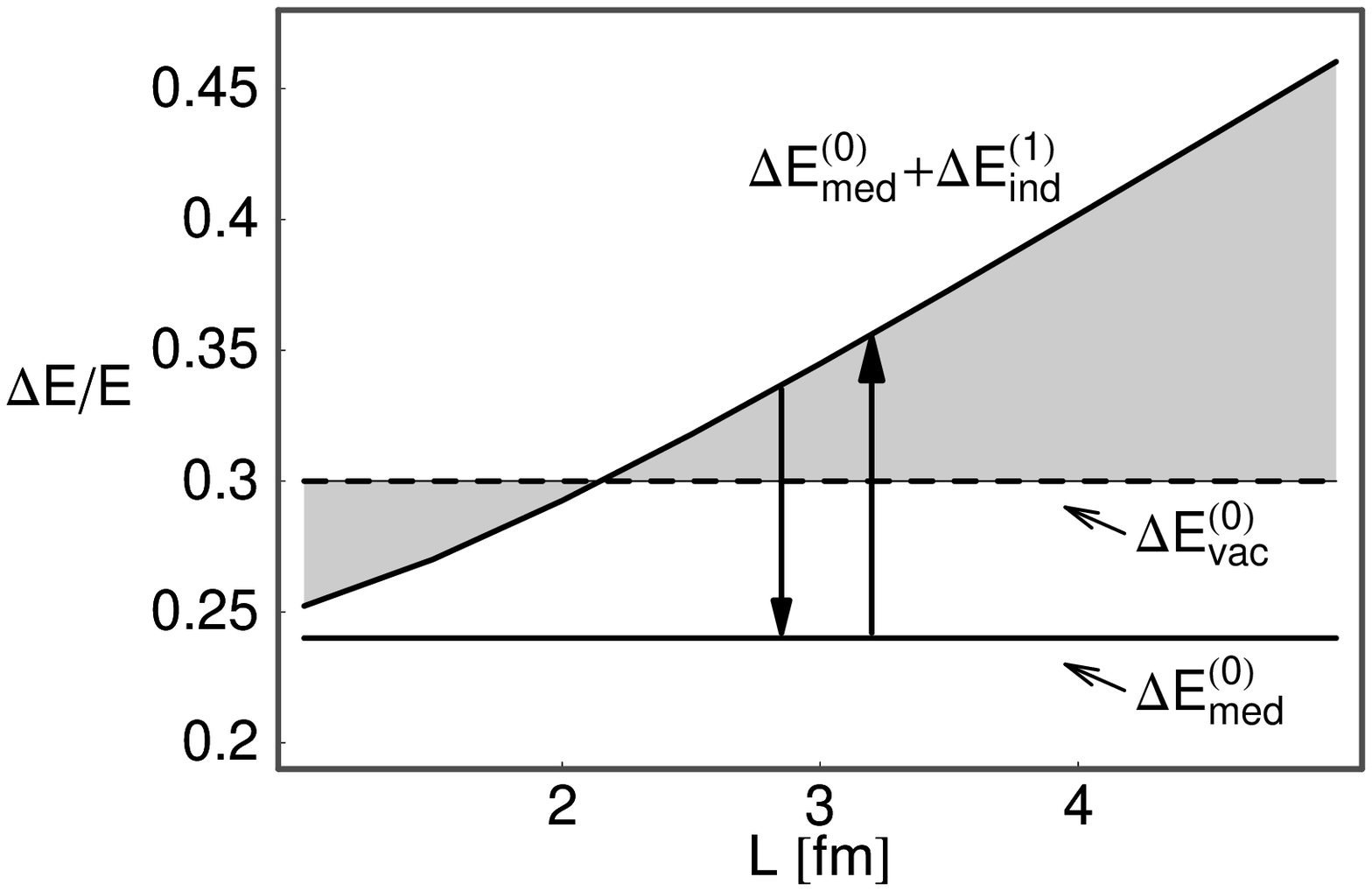}
\begin{minipage}[t]{15.0cm}
{\small {FIG~4.} The fractional energy loss for a 10 GeV charm quark is 
plotted versus the effective static thickness $L$ of a plasma characterized 
by $\mu=0.5$ GeV and $\lambda=1$ fm. The dashed middle horizontal line 
corresponds to the energy loss in the vacuum taking into account the 
kinematic dead cone of radiation for heavy quarks~{\protect\cite{Dokshitzer:2001zm}}. The lower horizontal solid line shows our estimate of the reduction 
of the zeroth order energy loss due to the QCD analog of the Ter-Mikayelian
effect. The solid curve corresponds to the net energy loss, 
$\Delta E^{(0)}_{med} + \Delta E^{(1)}_{ind}$. }
\end{minipage}
\end{center}
\vskip 4truemm 

We see that the net energy loss, $\Delta E^{(0)}_{med}+\Delta E^{(1)}_{ind}$, 
is found to be even smaller than in~\cite{Djordjevic:2003be} due to the 
corrections made in Eq.~(\ref{E1_ind}). Also, notice that for the effective 
static opacity medium~\cite{Levai_opacity} with $L$ in the range of $3-4$ fm, 
$\mu\approx 0.5\; {\rm GeV}$ and $\lambda\approx 1$fm, the difference between 
the medium energy loss ($\Delta E^{(0)}_{med}+\Delta E^{(1)}_{ind}$) and 
naive vacuum value ($\Delta E^{(0)}_{vac}$) is small, i.e. between $5\%$ and 
$10\%$ of the initial energy of the quark. These results, therefore, suggest 
that in addition to the heavy quark dead cone effect, 
the apparent null effect observed for heavy quark energy loss via single 
electrons may in part be due to a further reduction of the leading order 
energy loss. The different dependence of these effects on the plasma 
thickness $L$ and on the transport properties, $\mu$ and $\lambda$, should 
make it possible to test this explanation by varying the beam energy, $A$, 
and centrality at RHIC and LHC energies.

\subsection{Heavy quark energy loss at LHC}

In this section we want to estimate the heavy quark energy loss at LHC. 
According to some estimates~\cite{qm01}, the density of plasma partons at LHC 
should be 3-4 times larger than the one at RHIC. Therefore the results bellow 
are calculated using $\lambda$ and $\mu$ based on the estimate given above.

The numerical results for the first order induced radiative energy loss are 
shown on Fig.5 for charm and bottom quarks. We assume here that 
$\alpha_s=0.3, \; \mu=0.7\;{\rm GeV}$, $\lambda=0.7$ fm and $L=4$ fm for the 
plasma parameters. As before, the Ter-Mikayelian effect does not have 
significant effect on the induced energy loss. On the other hand, we see that 
the induced energy loss fraction is two times larger than before, i.e. 
$\Delta E^{(1)}/E \approx 0.3$ for charm quarks and about half for bottom.

\begin{center}
\vspace*{6.8cm} \includegraphics{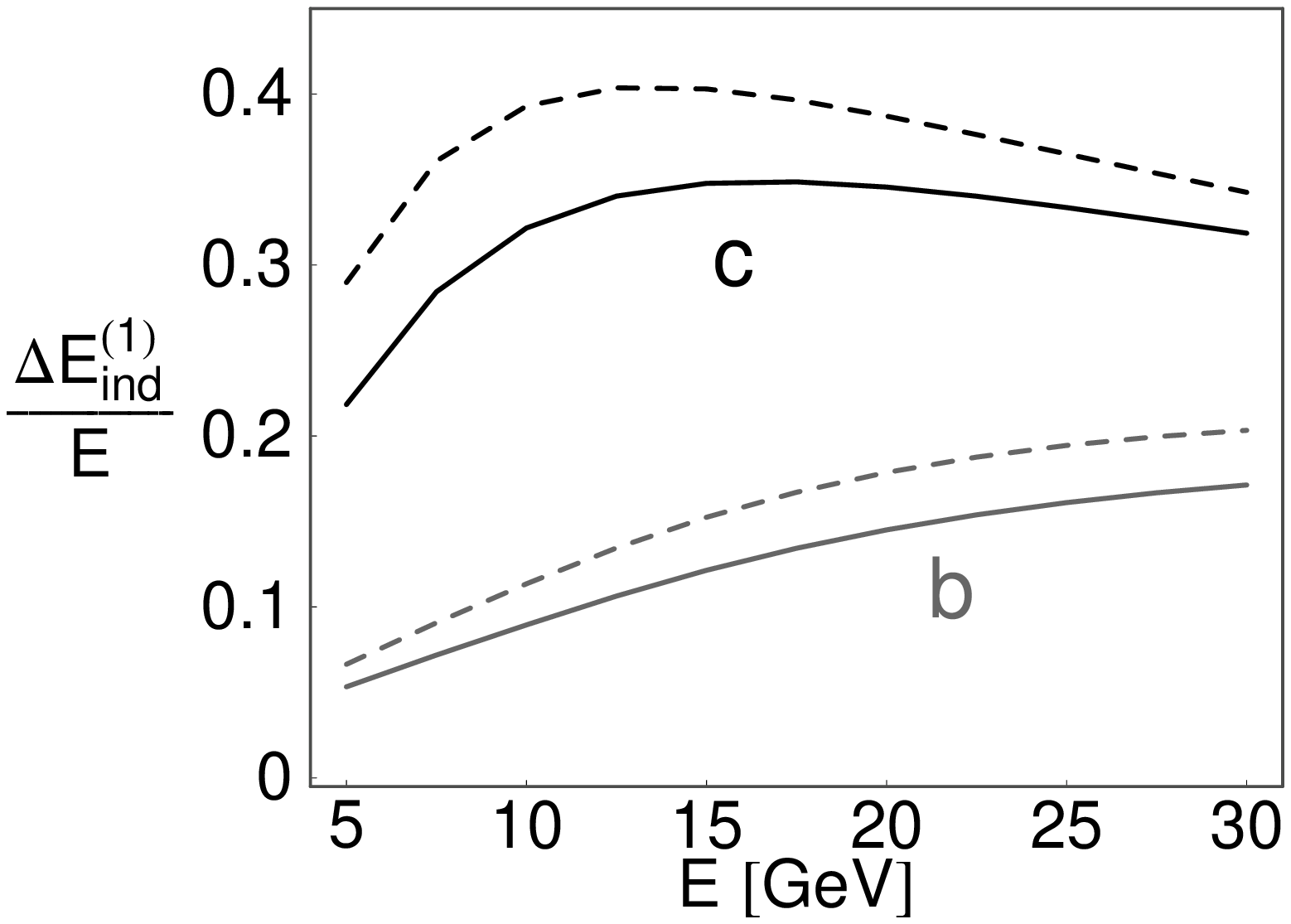}
\begin{minipage}[t]{15.0cm}
{\small {FIG~5.} The $1^{st}$ order in opacity fractional energy loss for 
heavy quarks with (solid curves) and without (dashed curves) Ter-Mikayelian 
effect approximated by a constant $m_g=\mu/\sqrt{2}$ for transverse modes 
only. Upper curves correspond to charm, and lower to bottom quarks as a 
function of their energy in a plasma characterized by $\alpha_s=0.3, \; 
\mu=0.7\;{\rm GeV}$, $\lambda=0.7$ fm and $L=4$ fm. }
\end{minipage}
\end{center}
\vskip 4truemm 

Fig.~6 shows the first order induced radiative energy loss for charm and 
bottom quarks as a function of opacity. We see that, again, for bottom quarks 
the thickness dependence is still closer to the linear Bethe-Heitler like 
form, $L^1$, than the asymptotic energy quadratic form~\cite{ELOSS,GLV}. 
However, for charm quark, the thickness dependence is somewhere between 
linear and quadratic form.

\begin{center}
\vspace*{6.7cm} \includegraphics{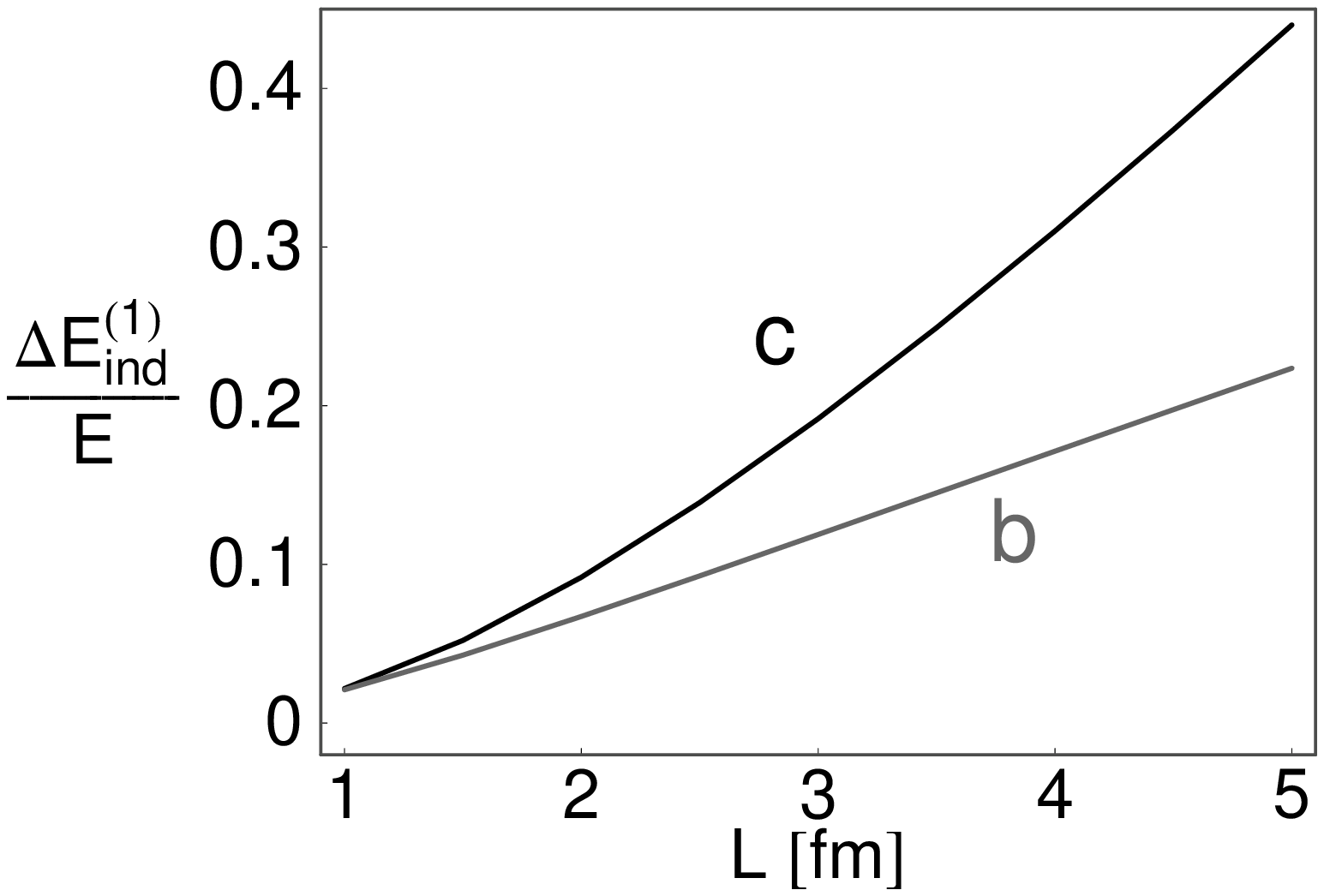}
\begin{minipage}[t]{15.0cm}
{\small {FIG~6.} The $1^{st}$ order in opacity fractional energy loss for a 
30 GeV charm and bottom quark is plotted versus the effective static 
thickness $L$ of a plasma characterized by $\mu=0.7$ GeV and $\lambda=0.7$ fm. 
Upper curve correspond to charm, and lower to bottom quark.}
\end{minipage}
\end{center}
\vskip 4truemm 

\begin{center}
\vspace*{6.8cm} \includegraphics{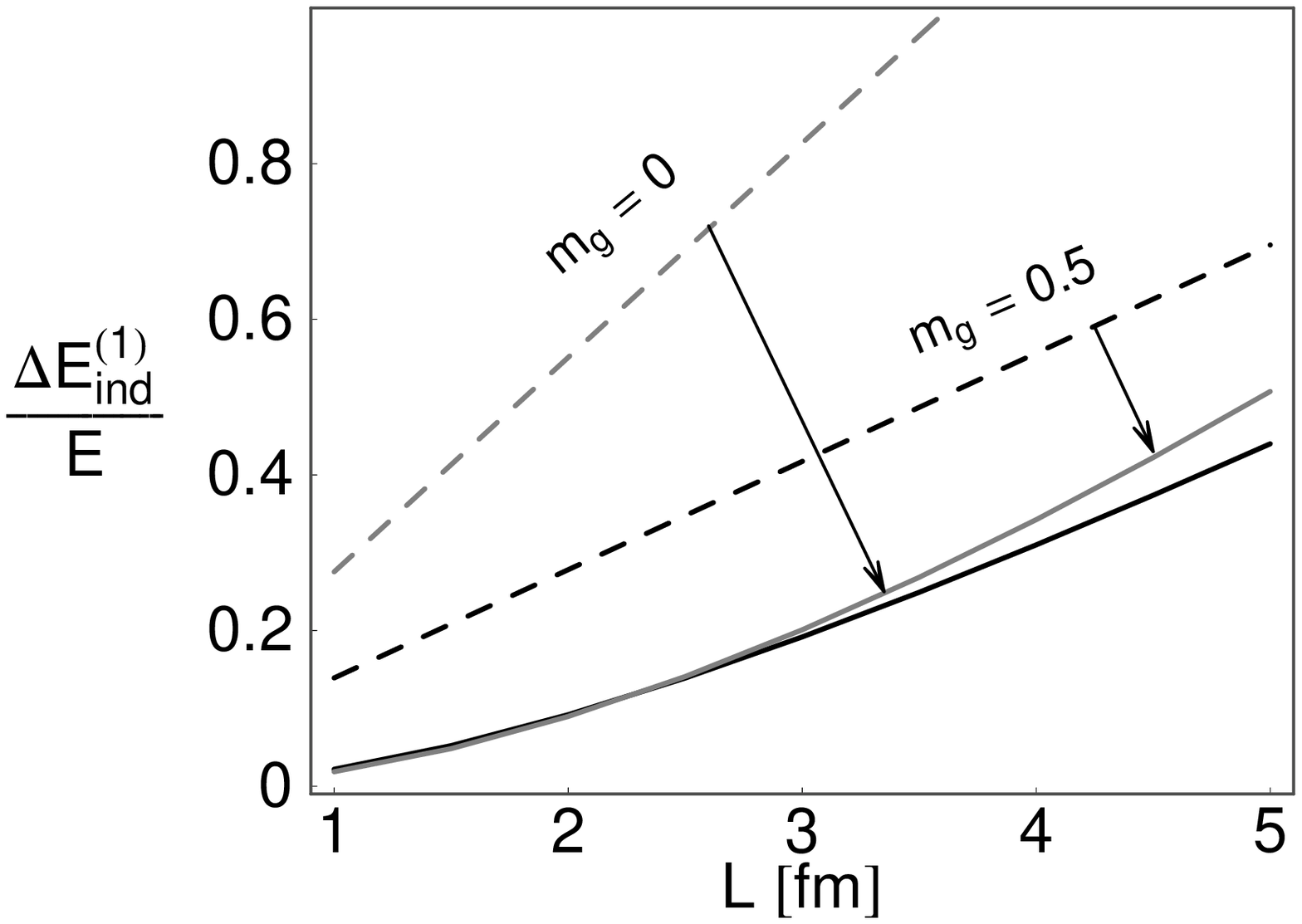}
\begin{minipage}[t]{15.0cm}
{\small {FIG~7.} Comparison of the incoherent limit (dashed 
curves) to the $1^{st}$ order (solid curves) in opacity fractional energy 
loss for $E=30$ GeV charm quarks  via Eq.(\protect{\ref{E1_ind}}) under 
estimated LHC conditions ($\alpha_s=0.3, \; \mu=0.7\;{\rm GeV}$). 
As in Fig. 3, the higher dashed and solid curves correspond to $m_g=0$,
while the lower to $m_g=\mu/\sqrt{2}$. Arrows point between corresponding 
incoherent limit  and full $1^{st}$ order results. }
\end{minipage}
\end{center}

Comparing Fig.7 to Fig. 3, we see that for $30$ GeV quarks, the first order 
result is now further away from the incoherent limit. The destructive LPM 
finite formation effects included in the full $1^{st}$ order result substantially
reduce the incoherent energy loss.

For LHC conditions, we can again, by using the results for associated energy 
loss from~\cite{Djordjevic:2003be}, compare the net 
($\Delta E^{(1)}_{ind}+\Delta E^{(0)}_{med}$) energy loss in the medium, with 
the one in the vacuum defined 
by $\mu_{vac}=0$ GeV, as shown on Fig.~8. We see that the dielectric plasmon 
effect reduces the $\Delta E^{(0)}_{med}/E$ form $\approx 1/3$ to 
$\approx 1/5$. As before, this contribution is independent of the 
thickness of the plasma as long as $L$ is not too small.

\begin{center}
\vspace*{6.8cm} \includegraphics{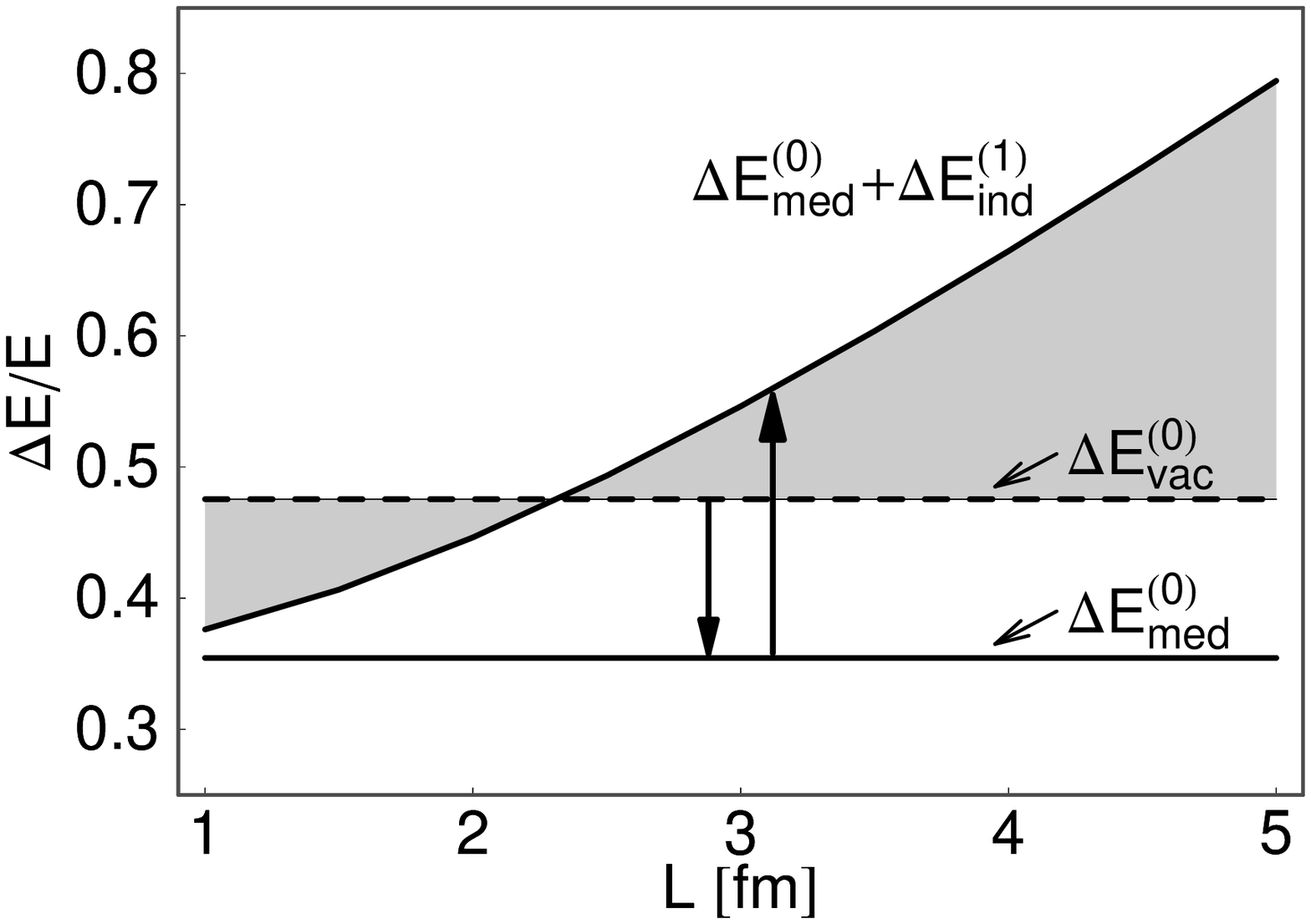}
\begin{minipage}[t]{15.0cm}
{\small {FIG~8.} The fractional energy loss for a 30 GeV charm quark is 
plotted versus the effective static thickness $L$ of a plasma characterized 
by $\mu=0.7$ GeV and $\lambda=0.7$ fm. The dashed middle horizontal line 
corresponds to the energy loss in the vacuum taking into account the 
kinematic dead cone of radiation for heavy quarks~{\protect\cite{Dokshitzer:2001zm}}. The lower horizontal solid line shows our estimate of the reduction 
of the zeroth order energy loss due to the QCD analog of the Ter-Mikayelian
effect. The solid curve corresponds to the net energy loss, 
$\Delta E^{(0)}_{med} + \Delta E^{(1)}_{ind}$. }
\end{minipage}
\end{center}
\vskip 4truemm 

We see that the difference between net energy loss, $\Delta E^{(0)}_{med}+
\Delta E^{(1)}_{ind}$ and naive vacuum value $\Delta E^{(0)}_{vac}$ at LHC is 
two times lager than the same difference at RHIC. Therefore, these results 
suggest a somewhat larger suppression of charm quark at 
LHC compared to RHIC.

\section{Summary}

In this paper we generalized the GLV radiative  energy loss formalism 
to heavy quarks including also the plasmon effects. We have shown that the 
reaction operator method introduced in~\cite{GLV} can also be applied to the 
case of massive quarks and gluons. Remarkably, a simple mass dependent energy 
shift, via Eqs.(\ref{omegm},\ref{hbgcdef}), was found to modify all direct 
and virtual diagrams. This result, proven in the appendices A-G made it 
possible to generalize the GLV zero mass results to the case of heavy quarks 
to all orders in opacity. We also derived the incoherent limit of the induced 
radiation spectrum from heavy quarks and found a closed form expression 
summed to all orders in opacity.

The dependence of the charm first order in opacity energy loss on the jet 
energy and plasma parameters were studied numerically.  We showed that the 
Ter-Mikayelian effect on the induced energy loss is comparable but somewhat 
larger than in~\cite{Dokshitzer:2001zm}. In addition, it was shown shown that 
the induced contribution increases approximately linearly with $L$, as first 
reported in~\cite{Djordjevic:2003be}. For charm quarks at RHIC the heavy 
quarks energy loss is close in magnitude and thickness dependence to the 
incoherent linear Bethe-Heithler like limit. This is in contrast to the 
quadratic (BDMS) thickness dependence characteristic for light quarks in the 
deep LPM regime. We note that ref.~\cite{Zhang:2003wk} reached similar 
conclusions using a different approach.

At LHC the partial LPM reduction of charm quark energy loss below the 
incoherent limit is predicted for $E=30$ GeV jets. However, the predictions 
for LHC are only quantitatively, but not qualitatively different from those 
at RHIC as long as the densities, and hence plasma parameters at LHC are not 
much more than about four times that at RHIC. 

Finally we show in Fig. 9, the predicted radiation length (see 
Eq.~(\ref{Lrad})) as a function of the heavy quark mass under estimated  RHIC 
and LHC quark gluon plasma conditions. The main feature to note is that the 
radiation length is comparable or greater than nuclear radii in all cases. 
The results imply that jet quenching is dominantly a volume emission rather 
than a surface emission phenomenon for all mass jets.
\begin{center}
\vspace*{6.9cm} \includegraphics{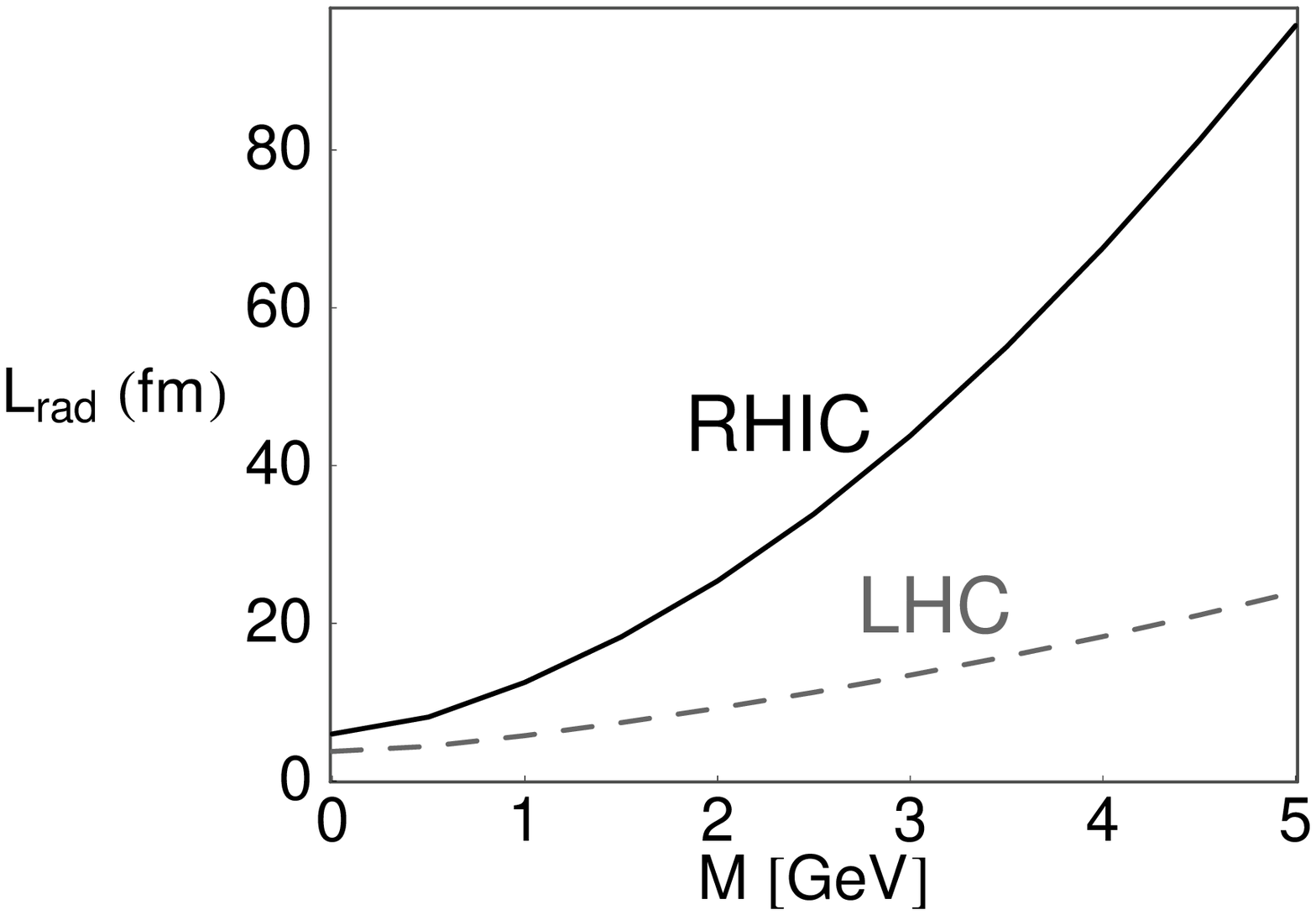}
\begin{minipage}[t]{15.0cm}
{\small {FIG~9.} The radiation length of charm quarks as a function
of their mass. Upper (lower) curve corresponds to the 10 (30) GeV charm quark 
at RHIC (LHC) conditions. }
\end{minipage}
\end{center}
\vskip 4truemm 
For light parton jets, the observed $R_{AA}\sim 0.2$ accidentally suggest
that only jets produced near the surface may survive to the detector.
However, as shown in ref.\cite{Vitev:2002pf}, the finite GLV energy loss
for jets propagating through the entire  volume of the plasma quantitatively
explains not only the magnitude and $p_T$ dependence of $R_{AA}(p_T)$ in 
central $Au+Au$ but also its centrality dependence~\cite{AGV}.
Heavy quark tomography will give an important  complementary  test of this 
physical picture. The computation presented here provides the basis for 
testing this theory on future experiments. The applications of heavy quark 
tomography to RHIC and LHC experiments is the subject of our current 
work.\\[2ex]

\vspace{1.0 cm}
{\em Acknowledgments:} Valuable discussions with I. Vitev, Z. Lin, J.
Nagle, X.N. Wang, and W. Zajc on heavy quark production at RHIC are gratefully
acknowledged. This work is supported by the Director, Office of
Science, Office of High Energy and Nuclear Physics, Division of
Nuclear Physics, of the U.S. Department of Energy under Grant No.
DE-FG02-93ER40764.

\begin{appendix}

\section{Apendix: Assumptions}

To compute the medium induced energy loss, we use the same assumptions 
as in~\cite{GLV}. For convenience,  we recall here those assumptions.

First, we consider a Yukawa potential as in Eqs.~(\ref{gwmod}) and 
assume that all the $x_j$ are distributed with the same density 
\beqar
\rho(\vec{\bf x})=\frac{N}{A_\perp} \bar{\rho}(z)
\;\; , 
\eeqar{densit}
where $\int dz\bar{\rho}(z)=1$.

Second,  we assume that the observed $p=[E^{+},E^{-},0]$  is 
high as compared to the potential screening scale, i.e.

\beqar
E^{+} \gg \mu \;\; .
\eeqar{assume2}

We also assume that the distance between the source and scattering
centers are  large compared to the interaction range

\beqar
z_i-z_0 \gg 1/\mu \;\; .
\eeqar{assume3}

Finally, we assume that the source current or packet $J(p)$ varies slowly 
over the range of momentum transfers supplied by the potential. 

A major simplification occurs if the relative transverse coordinate (impact 
parameter) ${\bf b}={\bf x}_{i }-{\bf x}_{0 }$  varies over a large 
transverse area, $A_\perp$, relative to the interaction area $1/\mu^2$. 
In this case,  the ensemble average over the scattering center
location reduces to an impact parameter average as follows

\beqar
\langle \, \cdots \, \rangle = \int \frac{d^2{\bf b}}{A_\perp} \cdots
\;\;.
\eeqar{impact}

The ensemble average over the phase factor then yields

\beqar
\langle \, e^{-i({\bf q}-{\bf q}^{\prime})\cdot
{\bf b}} \, \rangle= \frac{(2\pi)^2}{A_\perp} 
\delta^2({\bf q}-{\bf q}^{\prime})
\eeqar{bave}

Also, to calculate the diagrams with one and two scattering centers we need 
to find the mass correction for the full triple gluon vertices $\Lambda_{i}$,
and  $\Lambda_{ij}$ given in~\cite{GLV}. However, it is easy to show that in 
the approximation $M^{2}/p_{z}^{2}\ll 1$, these vertices remain the same. 
Therefore, the full triple gluon vertices including coupling and color algebra 
for producing a final color $c$ from initial color $a$ followed by color 
potential interactions $a_n$ and $a_m$ are then given by

\beqar
\Lambda_m &\equiv& \Gamma_m (-f^{ca_m a})(ig_st_a)(T_{a_m}(m))
\nonumber \\[1ex] 
&\approx&   \; -2g_s E^+{ \bf{\epsilon}}
\cdot({\bf k}-{\bf q}_{m}) [c,a_m] T_{a_m}(m) \;\;,  \nonumber \\[1ex]
\Lambda_{mn}&\equiv&  \Gamma_{mn}
(-f^{ca_n e})(-f^{ea_m a})(ig_st_a) (T_{a_n}(n))(T_{a_m}(m))
\nonumber \\[1ex]
&\approx& 
-2ig_s E^+ k^+{  \bf{\epsilon}}
\cdot({\bf k}-{\bf q}_{1}-{\bf q}_{2}) [[c,a_n],a_m] 
(T_{a_n}(n)T_{a_m}(m)) \;\;.
\eeqar{lam2}

\section{Diagrams $M_{1,0,0}$,   $M_{1,1,0}$  and   $M_{1,0,1}$  }

In this appendix we present explicit calculation of the diagrams shown in 
Fig. 8.

\begin{center}
\vspace*{5.5cm}
\includegraphics{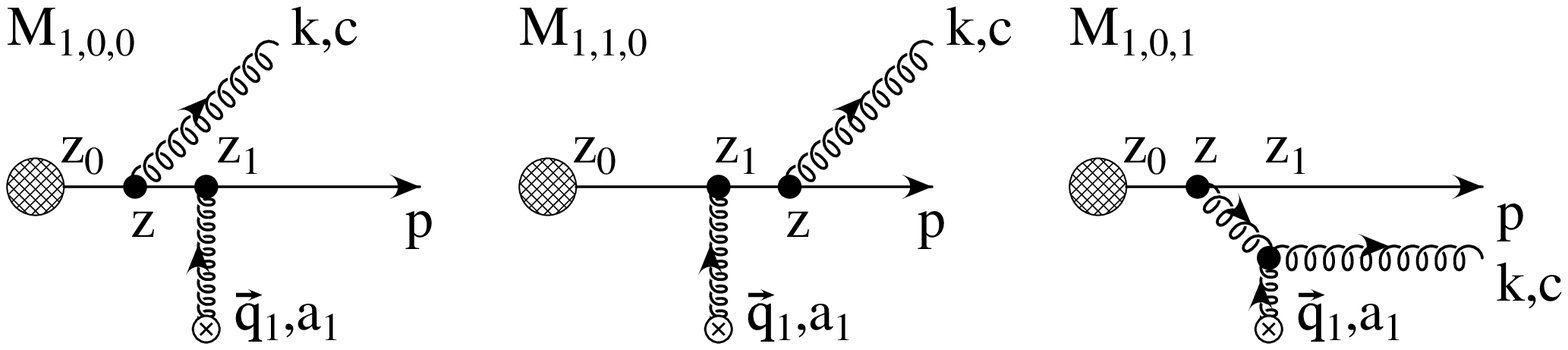}  
\vskip -20pt
\begin{minipage}[t]{15.0cm}
{\small {FIG~8.}
Three ``direct'' terms 
 $M_{1,0,0}$,   $M_{1,1,0}$  and   $M_{1,0,1}$
contribute to the soft gluon radiation amplitude 
to first order in opacity $  L/\lambda \propto \sigma_{el}/A_\perp$.}
\end{minipage}
\end{center}
\vskip 7truemm

\subsection*{Computation of $M_{1,0,1}$}

As a first application, consider the one rescattering
amplitude $M_{1,0,1}$. 

\beqar
M_{1,0,1} &=& \,\int\frac{d^4 q_1}{(2\pi)^4} \; iJ(p+k-q_1)
e^{i(p+k-q_1)x_0}\,\Lambda_1(p,k,q_1)V(q_1)e^{iq_1x_1} 
\, \times \nonumber \\[1ex]
&\;& \times \; i\Delta_{M}(p+k-q_1)(-i)\Delta_{m_{g}}(k-q_1)  
\nonumber \\[1ex]
& \approx&  J(p+k)
e^{i(p+k)x_0} \;[c,a_1] T_{a_1} \,(-i)\int
 \frac{d^2 {\bf q}_{1}}{(2\pi)^2}
\,e^{-i{\bf q}_{1} \cdot ({\bf x}_{ 1}-{\bf x}_{ 0})}
\, 2g_s \, { {\bf \epsilon} 
\cdot({\bf k}-{\bf q}_{1})} \, \times
\nonumber \\[1ex]
&\;& \times \; 2E  \, \int \frac{dq_{1z}}{2\pi}v(q_{1z},{\bf q}_{1}) 
\Delta_{M}(p+k-q_1)\Delta_{m_{g}}(k-q_1)
\,e^{-iq_{1z}(z_1-z_0)}  \;\; .
\eeqar{101}

The longitudinal momentum transfer integral, 

\beqar
I_1(p,k,{\bf q}_{1},z_1-z_0) \equiv
 \, \int \frac{dq_{1z}}{2\pi}v(q_{1z},{\bf q}_{1}) 
\Delta_{M}(p+k-q_1)\Delta_{m_{g}}(k-q_1) \,e^{-iq_{1z}(z_1-z_0)}  \;\;, 
\eeqar{int1}

can be performed via
closing the contour below the real axis since $z_1 > z_0$.

In addition to the potential singularities at $\pm i\mu_{1}$ 
($\mu^2_i\equiv \mu^2_{i\perp}={\bf q}_i^2+\mu^2 $),
the two propagators have two poles in the lower $q_{1z}$ plane,
which are approximately located at

\beqar
\begin{array}{ll}
\bar{q}_{1}=-\omega_0-\tilde{\omega}_{m}-i\epsilon\;\;,    \\[1.ex] 
\bar{q}_{2}=-\omega_0 + \omega_1 -i\epsilon\;\;. 
\end{array}
\eeqar{res20}

where $\omega_{0}=\frac{{\bf k}^{2}}{2\omega}$, $\omega_{i}=
\frac{({\bf k}-{\bf q}_{i})^{2}}{2\omega}$ and 
$\tilde{\omega}_{m}=\frac{m_{g}^{2}+M^{2}x^{2}}{2\omega}$. Note that 
$\tilde{\omega}_{m}$ is the term which has the information on finite mass 
corrections.

\medskip
 
The residues give
\beqar
{\rm Res}(\bar{q}_{1})\approx-v(-\omega_0-\tilde{\omega}_{m},{\bf q}_{1})
\frac{e^{i(\omega_0+\tilde{\omega}_{m})(z_1-z_0)}}{E^+k^+(\omega_1+\tilde{\omega}_{m})}\;,\nonumber \\[1ex] 
{\rm Res}(\bar{q}_{2})\approx 
v(\omega_1-\omega_0,{\bf q}_{1})\frac{e^{i (\omega_0 -\omega_1 )
(z_1-z_0)}}{E^+k^+(\omega_1+\tilde{\omega}_{m})}\;\;, 
\eeqar{resid1}

while

\beqar
{\rm Res}(-i\mu_{1})\approx\frac{4\pi \alpha_s \, 
e^{-\mu_{1}(z_1-z_0)}}
{(-2i\mu_{1}) E^+k^+(-i\mu_{1})^2} \;\;,
\eeqar{resid2}

where we assumed that $k^+\gg\mu_{1}\gg \omega_i$.

In the well-separated case where $\mu(z_1-z_0) = \mu \lambda\gg 1$,
this potential residue is exponentially suppressed
and therefore 

\beqar
I_1(p,k,{\bf q}_{1},z_1-z_0)&\approx& 
\frac{i}{E^+k^+(\omega_1+\tilde{\omega}_{m})}  \, \times
\nonumber \\[1ex]
&\;& \times \; \left(v(-\omega_0-\tilde{\omega}_{m},{\bf q}_{1})e^{i(\omega_0+\tilde{\omega}_{m})
(z_1-z_0)}- v(\omega_1-\omega_0,{\bf q}_{1}) e^{i (\omega_0 -\omega_1 ) (z_1-z_0)} \right)
\nonumber \\[1ex]
&\approx&  v(0,{\bf q}_{1})\frac{i}
{E^+k^+(\omega_1+\tilde{\omega}_{m})}
\left(e^{i(\omega_0+\tilde{\omega}_{m})(z_1-z_0)}-e^{i (\omega_0 -\omega_1 )(z_1-z_0)}\right) \;\;.
\eeqar{int1a}

Using the fact that 

\beqar
\omega_0+\tilde{\omega}_{m}=\frac{{\bf k}^{2}+m^{2}_{g}+M^{2}x^{2}}{2\omega}, \nonumber \\[1ex] 
\omega_0 -\omega_1 = \frac{{\bf k}^{2}-({\bf k-q}_{1})^{2}}{2\omega}
\eeqar{relations}

we finally get

 \beqar
&&M_{1,0,1} = J(p)e^{i(p+k)x_0} 
 \, (-i) \int \frac{d^2 {\bf q}_{1}}{(2\pi)^2}  v(0,{\bf q}_{1})
\,e^{-i{\bf q}_{1} \cdot {\bf b}_1} 2ig_s\,\frac{ { {\bf \epsilon} 
\cdot({\bf k}-{\bf q}_{1})}}
{({\bf k}-{\bf q}_1 \,)^2+M^{2}x^{2}+m^{2}_{g}} \;  
\; \times \nonumber \\[1ex]
&\;&  \qquad \qquad \times \; 
(e^{\frac{i}{2\omega}({\bf k}^{2}+m^{2}_{g}+M^{2}x^{2})(z_{1}-z_{0})}-
e^{\frac{i}{2\omega}({\bf k}^{2}-({\bf k-q}_{1})^{2})(z_{1}-z_{0})} ) \; [c,a_1] T_{a_1}\;\; ,
\eeqar{101b} 
where ${\bf b}_1={\bf x}_{1}-{\bf x}_{0}$.

The Eq.~(\ref{101b}) represents the massive corrections of Eq.~(A7) 
in~\cite{GLV}.

\subsection*{Computation of $M_{1,0,0}$}

Using the same technique as in the previous section we can now compute $M_{1,0,0}$.

\beqar
M_{1,0,0} &=& \,\int\frac{d^4 q_1}{(2\pi)^4} \; iJ(p+k-q_1)
e^{i(p+k-q_1)x_0}\,\,(ig_{s}) \epsilon_{ \alpha }
(2p-2q_{1}+k)^{\alpha} \times \nonumber \\[1ex]
&\;& \times \; i\Delta_{M}(p-q_1+k)i\Delta_{M}(p-q_1) (2p-q_{1})^{0} V(q_1)e^{iq_1x_1} T_{a_{1}} a_{1} c \nonumber \\[1ex]
& \approx&  J(p+k) e^{i(p+k)x_{0}} (-i g_{s} a_{1} c T_{a_{1}}) 2E 
\int \frac{d^{2}\mathbf{q}_{1}}{(2\pi)^{2}}
e^{ -i \mathbf{q}_{1} \mathbf{b}_{1}} I_{2},
\eeqar{100}

where

\beq
I_{2}(p,k,{\bf q}_{1},z_1-z_0) = \int \frac{dq_{z1}}{ 2 \pi} 
\frac{ \epsilon_{\alpha} (2p-2q_{1}+k)^{\alpha}} 
{(p-q_{1}+k)^{2} - M^{2} + i\epsilon}
\frac{1}{(p-q_{1})^{2} - M^{2} + i\epsilon} 
v( \vec{\mathbf{q}}_{1}) e^{-iq_{1z}(z_{1}-z_{0})}.
\eeq{}

Since $z_{1}>z_{0}$, integral $I_{2}$ can be performed via closing the 
contour below the real axis. Only roots $( - \omega_{0} - \tilde{\omega}_{m} -i 
\epsilon)$ and $( \frac{\mathbf{q}_{1}^{2}}{2p_{z}} - i\epsilon)$ contribute 
to this integral (root $-i\mu$ is suppressed, since $\mu (z_{1}-z_{0} ) 
\gg 1$). Using the relations given in the Eq.~(\ref{relations}) this integral 
becomes: 

\beq
I_{2}(p,k,{\bf q}_{1},z_1-z_0) =\frac{i}{E} \frac{ (\mathbf{ \epsilon \cdot k)} }{\mathbf{k}^{2} + m_{g}^{2}+M^{2}x^{2}} v( 0,\mathbf{q}_{1}) ( e^{\frac{i}{2\omega } 
(\mathbf{k}^{2} + m_{g}^{2}+M^{2}x^{2})(z_{1}-z_{0})}-1)
\eeq{}

Finally $M_{1,0,0}$ becomes

\beqar
&&M_{1,0,0} = J(p)e^{i(p+k)x_0} 
\, (-i)\int \frac{d^2 {\bf q}_{1}}{(2\pi)^2}  v(0,{\bf q}_{1})
\,e^{-i{\bf q}_{1} \cdot {\bf b}_1}  (2ig_s)\,\; \times \nonumber \\[1ex]
&\;& \times \; \frac{ { {\bf \epsilon} \cdot {\bf k}}}{\mathbf{k}^{2} + m_{g}^{2}+M^{2}x^{2}} \; ( e^{\frac{i}{2\omega } 
(\mathbf{k}^{2} + m_{g}^{2}+M^{2}x^{2})(z_{1}-z_{0})}-1) a_1 c T_{a_1}.
\eeqar{100final}

\subsection*{Computation of $M_{1,1,0}$}

\beqar
M_{1,1,0} &=& \,\int\frac{d^4 q_1}{(2\pi)^4} \; iJ(p+k-q_1)
e^{i(p+k-q_1)x_0} (2p+2k-q_{1})^{0}\,\,\times \nonumber \\[1ex]
&\;& \times \; i\Delta_{M}(p-q_1+k) i\Delta_{M}(p+k) (ig_{s}) \epsilon_{ \alpha } (2p+k)^{\alpha}  V(q_1)e^{iq_1x_1} T_{a_{1}} c a_{1} \nonumber \\[1ex]
& \approx&  J(p+k) e^{i(p+k)x_{0}} (-i g_{s} T_{a_{1}} c a_{1}) 
(2E+2\omega) 
\int \frac{d^{3}\mathbf{q}_{1}}{(2\pi)^{3}}
e^{ -i \vec{\mathbf{q}}_{1} ( \vec{ \mathbf{x}}_{1} - \vec{\mathbf{x}}_{0})} v( \vec{ \mathbf{q}}_{1}) \,\,\times \nonumber \\[1ex]
&\;& \times \; \frac{1}{(p - q_{1} + k)^{2} - M^{2} + i \epsilon} 
\frac{1}{(p+k)^{2} - M^{2} + i\epsilon} \epsilon_{\alpha} (2p+k)^{\alpha} \eeqar{110}

Since $ \omega \ll E \Longrightarrow ( 2E + 2 \omega) \approx 2E $.

\beq
\frac{ \epsilon_{\alpha} (2p+k)^{\alpha}} { (p+k)^{2} - M^{2} + i\epsilon} = 
\frac{2p \epsilon}{2pk+k^{2}}
\eeq{}

In the large $p_{z}$ limit this becomes

\beqar
\frac{ \epsilon_{\alpha } (2p+k)^{\alpha }}{(p+k)^{2} - M^{2} + 
i \epsilon } &=& 2 (1 + \frac{M^{2}}{4p_{z}^{2}}) 
\frac{ \mathbf{ \epsilon \cdot k}}{ \mathbf{k}^{2} + m_{g}^{2}+M^{2}x^{2}}
\approx 2 \frac{\mathbf{\epsilon \cdot k}}{\mathbf{k}^{2} + x^{2}M^{2} + 
m_{g}^{2}}
\eeqar{}

Using these two results, we get

\beq
M_{1,1,0} = J(p) e^{ipx_{0}} (-ig_{s}) T_{a_{1}} c a_{1} 4 E
\frac{\mathbf{\epsilon \cdot k}}{\mathbf{k}^{2} + m_{g}^{2}+M^{2}x^{2}}
\int \frac{d^{2} \mathbf{q}_{1}}{( 2\pi )^{2}} 
e^{-i \mathbf{q}_{1} \mathbf{b}_{1}} I_{3}(p,k,{\bf q}_{1},z_1-z_0)
\eeq{}

where

\beqar
I_{3}(p,k,{\bf q}_{1},z_1-z_0) &=& \int \frac{dq_{z1}}{2\pi} \frac{1}{(p-q_{1}+k)^{2}-M^{2}+i\epsilon}
e^{-iq_{1z}(z_{1}-z_{0})} v( \vec{ \mathbf{q}}_{1}) \;
\nonumber \\[1ex] &\approx&  (-i) \frac{ e^{i( \omega_{0} + \tilde{\omega}_{m})(z_{1}-z_{0})}}
{2p_{z}} v( 0, \mathbf{q}_{1}) 
\eeqar{I_3}

Therefore,

\beqar
&&M_{1,1,0} = J(p)e^{i(p+k)x_0} 
\, (-i)\int \frac{d^2 {\bf q}_{1}}{(2\pi)^2}  v(0,{\bf q}_{1})
\,e^{-i{\bf q}_{1} \cdot {\bf b}_1} \;  \, \times
\nonumber \\[1ex]
&\;& \times \; \; (-2ig_s)\,\frac{ { {\bf \epsilon} \cdot {\bf k}}}
{\mathbf{k}^{2} + m_{g}^{2}+M^{2}x^{2}} \;  e^{\frac{i}{2 \omega } 
(\mathbf{k}^{2} + m_{g}^{2}+M^{2}x^{2})(z_{1}-z_{0})}  \; c a_1 T_{a_1}\;\; ,
\eeqar{110final}

which in the massless limit, $M=m_{g}=0$, leads to Eq.~(A8) from~\cite{GLV}.

\section{Diagram $M_{2,0,3}$}

Consider next the gluon two-scattering  amplitude $M_{2,0,3}$. 
Fig.~9 shows that for inclusive processes two interesting cases arise.
\begin{center}
\vspace*{5.5cm}
\includegraphics{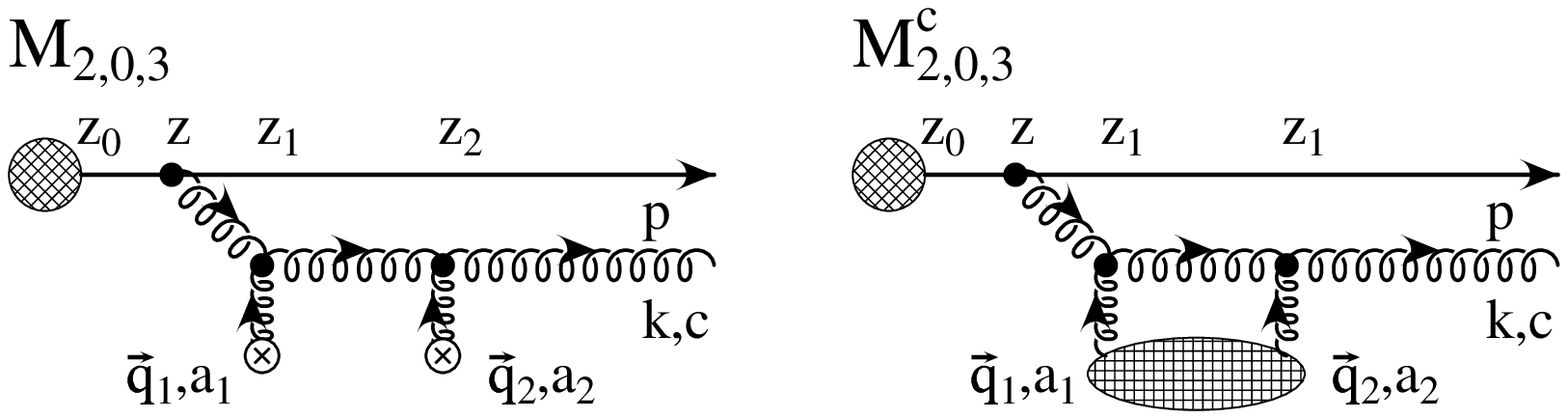}  
\vskip -15pt
\begin{minipage}[t]{15.0cm}
{\small { FIG~9.}
$M_{2,0,3}$  ``direct'' contributes to  
second order  in opacity 
$\propto (\sigma_{el}/A_\perp)^2$,
whereas $M^c_{2,0,3}=M_{2,0,3}(z_2=z_1)$ 
``contact-limit'' contribute to first order  in opacity 
$\propto (\sigma_{el}/A_\perp)^1$.}
\end{minipage}
\end{center}
\vskip 7truemm

In the Feynman diagram approach

\beqar
M_{2,0,3}&=& \, \int\frac{d^4 q_1}{(2\pi)^4}\frac{d^4 q_2}{(2\pi)^4}
\; iJ(p+k-q_1-q_2)e^{i(p+k-q_1-q_2)x_0} 
V(q_1)e^{iq_1x_1} V(q_2)e^{iq_2x_2} \, \times \nonumber\\[1.5ex]
&\;& \times \; \; \Lambda_{12}(p,k,q_1,q_2) \; i\Delta_{M}(p+k-q_1-q_2) 
(-i)\Delta_{m_{g}}(k-q_1-q_2)(-i)\Delta_{m_{g}}(k-q_2) \,
\nonumber \\[2ex]
&\approx&  J(p+k)e^{i(p+k)x_0}\; [[c,a_2],a_1] (T_{a_2}(2)T_{a_1}(1)) \,
\times \nonumber \\[1ex]
&\;& \times \; \; (-i)\int\frac{d^2 {\bf q}_{1}}{(2\pi)^2}
 (-i)\int \frac{d^2 {\bf q}_{2}}{(2\pi)^2}
\; 2ig_s {{\bf \epsilon}
\cdot({\bf k}-{\bf q}_{1}-{\bf q}_{2})} 
e^{-i{\bf q}_{1}\cdot{\bf b}_{1}} 
e^{-i{\bf q}_{2}\cdot{\bf b}_{2}}
\, \times \nonumber \\[1ex]
&\;& \times \int\frac{d q_{1z}}{(2\pi)}\frac{d q_{2z}}{(2\pi)}
\frac{(4E \omega)\, v(\vec{\bf q}_1)v(\vec{\bf q}_2)e^{-iq_{1z}(z_{1}-z_0)}
e^{-iq_{2z}(z_{2}-z_0)}} 
{((p+k-q_1-q_2)^2+M^{2}+i\epsilon)((k-q_1-q_2)^2+m^{2}_{g}+i\epsilon)
((k-q_2)^2+m^{2}_{g}+i\epsilon)} ,
\eeqar{203b}

where ${\bf b}_i={\bf x}_{i}-{\bf x}_{0}$ are transverse impact parameters,
and we used the soft gluon and rescattering kinematical simplifications
Eqs.~(\ref{lam2}), e.g. $J(p+k-q_1-q_2)\approx J(p+k)\approx J(p)$. For the 
$q_{1z}$ integral, it is convenient to  rewrite the phase as 

$ e^{-iq_{1z}(z_{1}-z_0)} e^{-iq_{2z}(z_{2}-z_0)}=
e^{-i(q_{1z}+q_{2z})(z_{1}-z_0)} e^{-iq_{2z}(z_{2}-z_1)}. $

The first longitudinal integral is closely related to Eq.~(\ref{int1})

\beqar
I_2(p,k,{\bf q}_{1}, \vec{\bf q}_2,z_1-z_0) = \int\frac{d q_{1z}}{2\pi}
\frac{ v(q_{1z},{\bf q}_{1})e^{-i(q_{1z}+q_{2z})(z_{1}-z_0)}}
{((p+k-q_1-q_2)^2+M^{2}+i\epsilon)((k-q_1-q_2)^2+m^{2}_{g}+i\epsilon)} \;\;.
\eeqar{i2}

Since $z_1-z_0\gg 1/\mu$,  we again close the contour  in the lower 
half $q_{1z}$ plane and neglect the pole at $- i\mu_{1}$.
The  remaining $q_{1z}$ poles are shifted by 
$-q_{2z}$ and ${\bf q}_{1}\rightarrow {\bf q}_{1}+{\bf q}_{2}$
relative to  Eq.~(\ref{res20}):

\beqar
\begin{array}{ll}
\bar{q}_{1}=-q_{2z}-\omega_0-\tilde{\omega}_{m}-i\epsilon\;,  \\[1.ex]
\bar{q}_{2}=-q_{2z}-\omega_0 +\omega_{(12)} -i\epsilon\;\;,
\end{array}
\eeqar{res21}

where $\omega_{(12)}= \frac{({\bf k}-{\bf q}_{1}-{\bf q}_{2})^2}{2\omega}.$
The residues at $\bar{q}_{1},\, \bar{q}_{2}$ then give

\beqar
I_2\approx \frac{i\left(v(-q_{2z}-\omega_0-\tilde{\omega}_{m},
{\bf q}_{1})e^{i(\omega_0+\tilde{\omega}_{m})(z_{1}-z_{0})} -
v(-q_{2z}-\omega_0+\omega_{(12)},{\bf q}_{1})
e^{i(\omega_{0}-\omega_{(12)})(z_1-z_0)} \right)}
{E^+k^+(\omega_{(12)}+\tilde{\omega}_{m})}
\;\; ,\eeqar{i2a}

where we have neglected ${\cal O} ({\rm exp}(-\mu \lambda))$ contributions. 
This differs from Eq.~(\ref{int1a}) mainly in that the potential is evaluated
near $-q_{2z}$, which still remains to be integrated over,
and $\omega_1\rightarrow \omega_{(12)}$.

Next we need the following  {\em critical} $q_{2z}$ integral

\beqar
I_3(k,{\bf q}_{1},{\bf q}_{2},z_2-z_1) &\equiv& 
\int\frac{d q_{2z}}{2\pi}
\frac{ v(-q_{2z}+\delta\omega,
{\bf q}_{1})v(q_{2z},{\bf q}_{2})e^{-iq_{2z}(z_2-z_1)}}
{((k-q_2)^2+m^{2}_{g}+i\epsilon)} \;\;.
\eeqar{i3a}

In the  general case  (including the special contact case with $z_2=z_1$)
both $q_{2z}=-i\mu_{2},\; -i\mu_{1}  $ singularities
in the Yukawa potential contribute together with the pole at 
$q_{2z}=\omega_2-\omega_0-i\epsilon$,
resulting in 
\beqar
I_3(k,{\bf q}_{1},{\bf q}_{2},z_2-z_1) &\approx&
\frac{-i}{k^+} \left[ \, v(0,{\bf q}_{1}) v(0,{\bf q}_{2})
\;e^{-i(\omega_2-\omega_0)(z_2-z_1)}  \right. \nonumber \\[1ex]
 &\;& \; \left.\;\;  - 
\frac{(4\pi \alpha_s)^2}{2\,(\mu_{1}^2-\mu_{2}^2)} \left(
\frac{e^{-\mu_{2} (z_2-z_1)}}{\mu_{2}^2}
-\frac{e^{-\mu_{1} (z_2-z_1)} e^{-i\delta\omega (z_2-z_1)}}
{\mu_{1}^2} \right)\, \right]
\;\; .
\eeqar{i3a2}

Fortunately, we are interested in only two extreme limits: 
\begin{itemize}
\item{The limit of  well-separated scattering centers $z_2-z_1\gg 1/\mu$} ;
\item{The special ``contact'' $z_2=z_1$ limit to compute unitary 
contributions.}
\end{itemize}

For $z_2-z_1=\lambda\gg 1/\mu$ the Eq.~(\ref{i3a2}) reduces to

\beqar
I_3(k,{\bf q}_{1},{\bf q}_{2},z_2-z_1\gg1/\mu) &\approx&
-\frac{i}{k^+}v(0,{\bf q}_{1}) v(0,{\bf q}_{2})
e^{-i(\omega_2-\omega_0)(z_2-z_1)}
\;\; .
\eeqar{i3a1}

For the special contact contribution $z_2-z_1=0$ it reduces to

\beqar
I_3(k,{\bf q}_{1},{\bf q}_{2},0) &\approx&
\frac{-i}{2\,k^+} v(0,{\bf q}_{1}) v(0,{\bf q}_{2})\;\; .
\eeqar{i3a3}

i.e., exactly $\half$ of the strength in Eq.~(\ref{i3a1}).

The contact limit of this amplitude is therefore

\beqar
M^{c}_{2,0,3}&\approx &  J(p)e^{i(p+k)x_0}
(-i)\int \frac{d^2 {\bf q}_{1}}{(2\pi)^2}
\, v(0,{\bf q}_{1})\,
e^{-i{\bf q}_{1}\cdot {\bf b}_{1}}
(-i)\int\frac{d^2 {\bf q}_{2}}{(2\pi)^2}
\, v(0,{\bf q}_{2})\,
e^{-i{\bf q}_{2}\cdot {\bf b}_{2}}
\; \times \nonumber \\[1ex]
&\;&  \times 
\; \frac{1}{2}\;(2ig_s)\;\frac{{ {\bf \epsilon}
\cdot({\bf k}-{\bf q}_{1}-{\bf q}_{2})}} 
{({\bf k}-{\bf q}_{1}-{\bf q}_{2})^2+M^{2}x^{2}+m^{2}_{g}}
\; 
\; [[c,a_2],a_1] (T_{a_2}T_{a_1})\;\; \times \nonumber \\[1ex]
&\;&  \times \{ e^{ \frac{i}{2\omega } ( \mathbf{k}^{2} + m_{g}^{2}+M^{2}x^{2})
(z_{1}-z_{0})} - e^{ \frac{i}{2\omega } ( \mathbf{k}^{2} - 
( \mathbf{k-q}_{1} - \mathbf{q}_{2})^{2}) (z_{1}-z_{0})} \}.
\eeqar{203d}

Note that in the massless limit it reduces to Eq.~(B10) from~\cite{GLV}.

\section{Diagrams $M_{2,0,0}$ and $M_{2,2,0}$}

In those graphs it is the jet rather than the gluon
that suffers two sequential scatterings as seen from Fig.~10.
\begin{center}
\vspace*{12.0cm}
\includegraphics{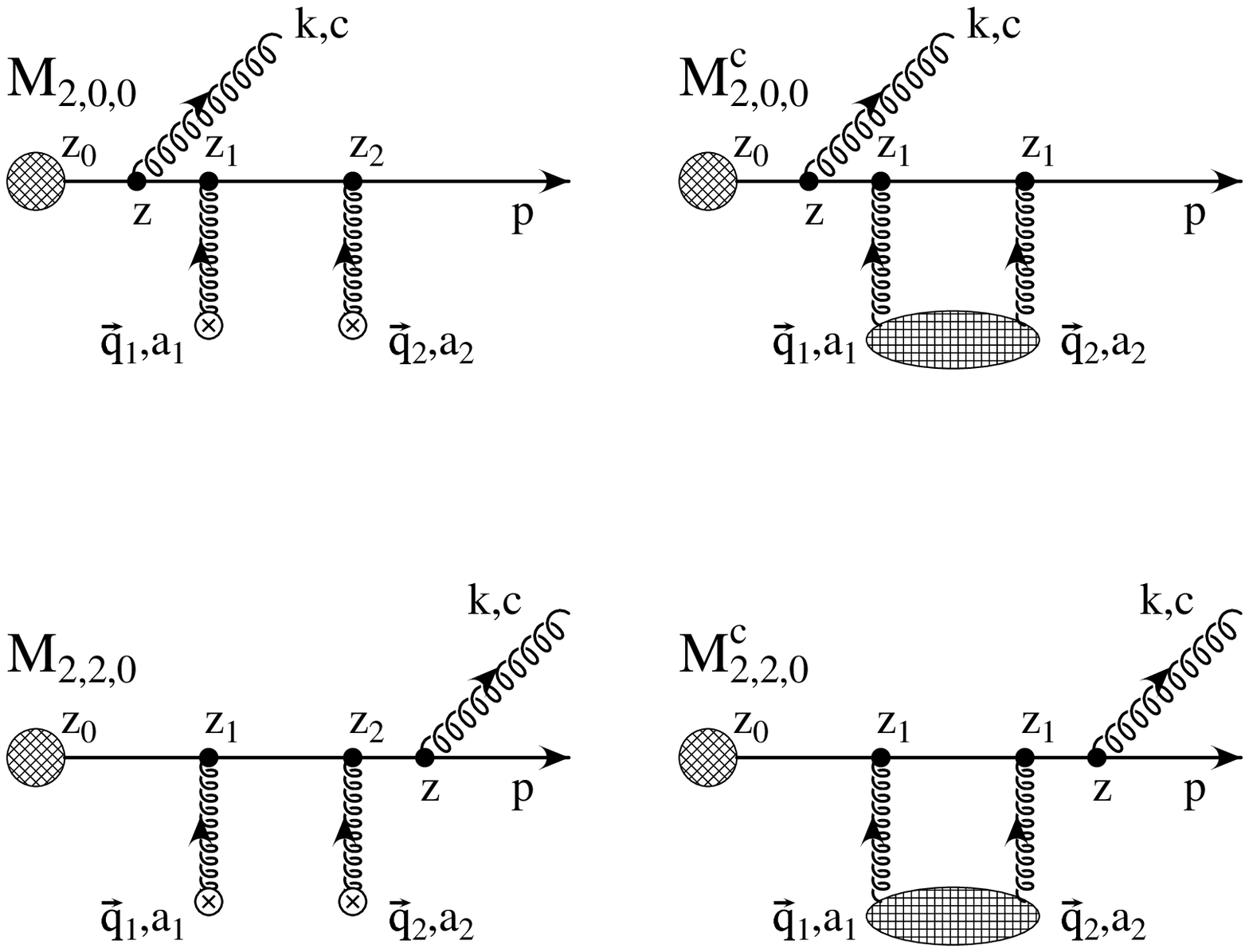}  
\vskip -30pt
\begin{minipage}[t]{15.0cm}
{\small { FIG~10.}
 $M_{2,0,0}$ and $M_{2,2,0}$  graphs in the well-separated case 
together with their $z_2=z_1$ limits  
$M^c_{2,0,0}$, $M^c_{2,2,0}$.}
\end{minipage}
\end{center}

\beqar
M_{2,0,0}&=&\int\frac{d^4 q_1}{(2\pi)^4}\frac{d^4 q_2}{(2\pi)^4}
\; iJ(p+k-q_1-q_2)e^{i(p+k-q_1-q_2)x_0} 
V(q_1)e^{iq_1x_1} V(q_2)e^{iq_2x_2} \; \times \nonumber\\[1ex]
&\;& \times \; (-i2E)^2ig_s(2p+k)_\mu \epsilon^\mu \; 
i\Delta_{M}(p+k-q_1-q_2) \,i\Delta_{M}(p-q_1-q_2)\,i\Delta_{M}(p-q_2) a_2 a_1 c(T_{a_2}T_{a_1}) \nonumber \\[2ex]
&\approx&  J(p)e^{ipx_0}(-i)\int\frac{d^2 {\bf q}_{1}}{(2\pi)^2}
e^{-i{\bf q}_{1}\cdot{\bf b}_{1}} 
(-i)\int \frac{d^2 {\bf q}_{2}}{(2\pi)^2}
e^{-i{\bf q}_{2}\cdot{\bf b}_{2}} \frac{2ig_s({\bf \epsilon}\cdot{\bf k})}{x} 
\; a_2a_1 c (T_{a_2}T_{a_1})\;(2 E)^2\
\; \times \nonumber \\[1.5ex]
&\;&\times \; \int\frac{d q_{1z}}{(2\pi)}\frac{d q_{2z}}{(2\pi)}
\frac{ v(\vec{\bf q}_1)v(\vec{\bf q}_2)e^{-iq_{1z}(z_{1}-z_0)}
e^{-iq_{2z}(z_{2}-z_0)}}
{((p+k-q_1-q_2)^2+M^{2}+i\epsilon)((p-q_1-q_2)^2+M^{2}+i\epsilon)
((p-q_2)^2+M^{2}+i\epsilon)} \;\;. 
\eeqar{200}

In this case we define

\beqar
I_2(p,k,{\bf q}_{1},\vec{\bf q}_2,z_1-z_0) = \int\frac{d q_{1z}}{2\pi}
\frac{ v(q_{1z},{\bf q}_{1})e^{-i(q_{1z}+q_{2z})(z_{1}-z_0)}}
{((p+k-q_1-q_2)^2+M^{2}+i\epsilon)((p-q_1-q_2)^2+M^{2}+i\epsilon)} \;\;.
\eeqar{i2prim}

Since $z_1-z_0\gg 1/\mu$, we  neglect the pole at $- i\mu_{1}$.
The  remaining $q_{1z}$ poles are 

\beqar
\begin{array}{ll}
\bar{q}_{1}=-q_{2z}-\omega_0-\tilde{\omega}_{m}-i\epsilon\;,        \\[1.5ex]
\bar{q}_{2}=-q_{2z}  -i\epsilon\;\;,
\end{array}
\eeqar{res2002}

where we discarded $({\bf p}+{\bf k} - 
{\bf q}_{1}-{\bf q}_{2})^2/E^+$
relative to $\omega_0$.  The $\bar{q}_{1},\, \bar{q}_{2}$ residues then give

\beqar
I_2\approx -i \frac{v(-q_{2z},{\bf q}_{1})}{(E^+)^2(\omega_{0}+\tilde{\omega}_{m})}
\left(e^{i(\omega_0+\tilde{\omega}_{m})(z_1-z_0)}-1\right) \;\;.
\;\; \eeqar{i2ab}

Not that $\omega_0$ has been neglected in the potential relative to
$\mu_{1}$. In the contact limit, the second integral, $I_3$, is then equal to

\beqar
\bar{I}_3(p,{\bf q}_{1},{\bf q}_{2},z_2-z_1) 
&\equiv& \int\frac{d q_{2z}}{(2\pi)}
\frac{ v(-q_{2z},{\bf q}_{1}) \, v(q_{2z},
{\bf q}_{2}) \, e^{-iq_{2z}(z_2-z_1)}}{((p-q_2)^2+M^{2}+i\epsilon)}
\nonumber \\[1ex]
&\approx & \frac{i}{E^+}v(0,{\bf q}_{1})v(0,{\bf q}_{2}) 
\, \times \, \left\{\begin{array}{ll}
1 \quad  &{\rm if } \;\; \mu\lambda = \mu(z_2-z_1) \rightarrow \infty \\[1.ex]
\frac{1}{2} \quad  & {\rm if } \;\; \mu\lambda = \mu(z_2-z_1) \rightarrow 0
\end{array} 
\right. \;.
\eeqar{i3200}

With the help of Eqs.~(\ref{i2ab},\ref{i3200})  in the case of contact limit 
we obtain

\beqar
M_{2,0,0}^{c}&=& \frac{1}{2} J(p)e^{i(p+k)x_0}\int\frac{d^2 {\bf q}_{1}}{(2\pi)^2} \frac{d^2 {\bf q}_{2}}{(2\pi)^2} v(0,{\bf q}_{1}) v(0,{\bf q}_{2})
e^{-i({\bf q}_{1}+{\bf q}_{2}) \cdot{\bf b}_{1}} \; \times \nonumber  \\[1ex]
&\;& \times \; (-2ig_s)\frac{({\bf \epsilon}
\cdot{\bf k})}{\mathbf{k}^{2} + m_{g}^{2}+M^{2}x^{2}} 
\; \{ e^{\frac{i}{2\omega } (\mathbf{k}^{2} + m_{g}^{2}+M^{2}x^{2})
(z_{1}-z_{0})}-1 \}
 \; a_2a_1 c\, (T_{a_2}T_{a_1}) \;\;.
\eeqar{200fin}

In the same way we will obtain for $M_{2,2,0}$

\beqar
M_{2,2,0}^{c}&=& \frac{1}{2} J(p)e^{i(p+k)x_0} \int\frac{d^2 {\bf q}_{1}}{(2\pi)^2} \frac{d^2 {\bf q}_{2}}{(2\pi)^2} v(0,{\bf q}_{1}) v(0,{\bf q}_{2})
e^{-i({\bf q}_{1}+{\bf q}_{2}) \cdot{\bf b}_{1}} \; \times \nonumber  \\[1ex]
&\;& \times \; (2ig_s) \frac{({\bf \epsilon}
\cdot{\bf k})}{\mathbf{k}^{2} + m_{g}^{2}+M^{2}x^{2}} 
\; e^{\frac{i}{2\omega } (\mathbf{k}^{2} + m_{g}^{2}+M^{2}x^{2})
(z_{1}-z_{0})}
 \; c a_2a_1 \, (T_{a_2}T_{a_1}) \;\;.
\eeqar{220fin}

\section{Diagrams $M_{2,0,1}$ and $M_{2,0,2}$}

Bellow we compute the case when one of the hits is on the parent parton and 
the other hit is on the radiated gluon. Explicit calculation is shown on the 
example of $M^c_{2,0,1} = M_{2,0,1}(z_2=z_1) $ in Fig.~11. 

\begin{center}
\vspace*{5.5cm}
\includegraphics{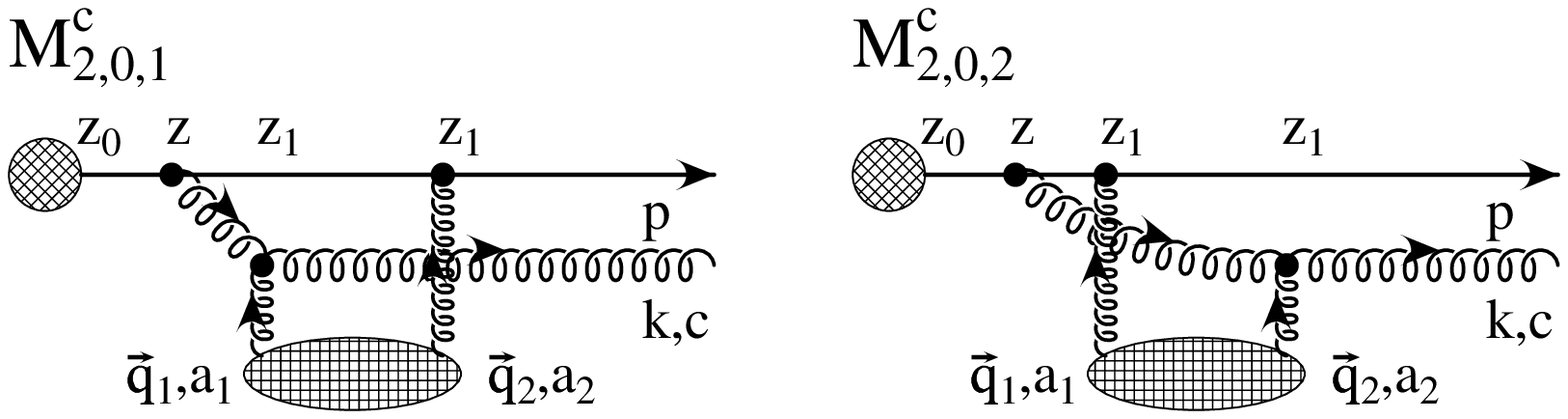}  
\vskip -10pt
\begin{minipage}[t]{15.0cm}
{\small { FIG~11.} $M^c_{2,0,1}$  and $M^c_{2,0,2}$ topologically 
indistinct   contact diagrams. There are no additional factors of $\half$ 
arising from the integration  in taking the $z_2= z_1$ limit in  $M_{2,0,1}$ 
and $M_{2,0,2}$.}
\end{minipage}
\end{center}

\beqar
&& \nonumber\\[1ex]
M_{2,0,1}&=&\int\frac{d^4 q_1}{(2\pi)^4}\frac{d^4 q_2}{(2\pi)^4}
\; iJ(p+k-q_1-q_2)e^{i(p+k-q_1-q_2)x_0} 
V(q_1)e^{iq_1x_1} V(q_2)e^{iq_2x_2} \; \times \nonumber\\[1ex]
&\;& \times \; (-iE^+) \Lambda_1 \; i\Delta_{M}(p+k-q_1-q_2) 
\,(-i)\Delta_{m_{g}}(k-q_1)\,i\Delta_{M}(p-q_2) \;
\nonumber \\[1ex]
&\approx&  J(p)e^{ipx_0}(-i)\int\frac{d^2 {\bf q}_{1}}{(2\pi)^2}
e^{-i{\bf q}_{1}\cdot{\bf b}_{1}} 
(-i)\int \frac{d^2 {\bf q}_{2}}{(2\pi)^2}
e^{-i{\bf q}_{2}\cdot{\bf b}_{2}}  \; \times \nonumber  \\[1ex]
&\;& \times \;2ig_s({\bf \epsilon}\cdot({\bf k}-{\bf q}_{1})) 
\; e^{i \omega_0 z_0} \; a_2 [c,a_1] (T_{a_2}T_{a_1})\;(2E)^2\
\; \times \nonumber \\[1.5ex]
&\;&\times \; \int\frac{d q_{1z}}{(2\pi)}\frac{d q_{2z}}{(2\pi)}
\frac{ v(\vec{\bf q}_1)v(\vec{\bf q}_2)e^{-iq_{1z}(z_{1}-z_0)}
e^{-iq_{2z}(z_{2}-z_0)}}
{((p+k-q_1-q_2)^2+M^{2}+i\epsilon)((k-q_1)^2+m^{2}_{g}+i\epsilon)
((p-q_2)^2+M^{2}+i\epsilon)} .  
\eeqar{201}

We perform the $q_{1z}$ integral first

\beqar
I_2(p,k, {\bf q}_{1},  \vec{\bf q}_2,z_1-z_0) = \int\frac{d q_{1z}}{2\pi}
\frac{ v(q_{1z},{\bf q}_{1})e^{-i(q_{1z}+q_{2z})(z_{1}-z_0)}}
{((p+k-q_1-q_2)^2+M^{2}+i\epsilon)((k-q_1)^2+m^{2}_{g}+i\epsilon)} \;\;.
\eeqar{201i2}

The pole at $-i\mu_{1}$ is again exponentially suppressed. 
The poles of interest in the lower half plane are 
 $q_{1z} = -q_{2z}-\omega_0-\tilde{\omega}_{m} - i\epsilon$ and
$q_{1z} = -\omega_0 + \omega_1 - i\epsilon$.

Taking the residues leaves us with

\beqar
&&I_2(p,k, {\bf q}_{1},  \vec{\bf q}_2,z_1-z_0) = \; \frac{i}
{E^+ k^+ (q_{2z}+\omega_1+\tilde{\omega}_{m})} \; \times \nonumber  \\[1ex]
&\;& \; \; \times \; \left(v(-q_{2z}-\omega_0 -\tilde{\omega}_{m}, {\bf q}_{1} ) e^{i(\omega_0+\tilde{\omega}_{m})(z_{1}-z_0)} -  v(\omega_1-\omega_0, {\bf q}_{1} )
e^{i (\omega_{0}-q_{2z} -\omega_{1}) (z_{1}-z_0)} \right) \;\;.  
\eeqar{i2res}

It is important to  notice that there is  no pole
at $q_{2z} = -\omega_1$ in Eq.~(\ref{i2res}). The
remaining integral over $q_{2z}$ is

\beqar
&&I_3(p,k,{\bf q}_{1},{\bf q}_{2},z_1-z_0,z_2-z_1 ) = \; \int 
\frac{d q_{2z}}{2\pi} \frac{1}{q_{2z}+\omega_1+
\tilde{\omega}_{m}} ( \frac{e^{-i(q_{2z}(z_{2}-z_1)-
(\omega_0+\tilde{\omega}_{m})(z_{1}-z_0))}}
{(p-q_2)^2+M^{2}+i\epsilon} \; \times \nonumber  \\[1ex]
&\;& \times \;  v(-q_{2z}-\omega_0-\tilde{\omega}_{m}
, {\bf q}_{1} )v(q_{2z}, {\bf q}_{2} ) )
 - \frac{e^{-i(q_{2z}(z_{2}-z_0)+(\omega_1-\omega_0)(z_{1}-z_0)  )}}
{(p-q_2)^2+M^{2}+i\epsilon} \; v(\omega_1 -\omega_0, {\bf q}_{1} )
v(q_{2z}, {\bf q}_{2} )).
\eeqar{201i3}

The poles in the lower half plane are $q_{2z}=-i\epsilon$,
$q_{2z}=-i\mu_{2}$, and  $q_{2z}=-\omega_0-\tilde{\omega}_{m}-i\mu_{1}$. In 
the well separated case, contributions from second and third residue are
exponentially suppressed $\propto \exp[-\mu_{2}(z_1-z_0)]$, and therefore 
can be neglected. In the contact limit contributions from second and third 
residue cancel exactly (for more details see Eq.~(d5)~\cite{GLV}. Therefore, 
we get

\beqar
&&M^c_{2,0,1}= J(p)e^{i(p+k)x_0}(-i)\int\frac{d^2 {\bf q}_{1}}{(2\pi)^2}
e^{-i{\bf q}_{1}\cdot{\bf b}_{1}}v(0,{\bf q}_{1})
(-i)\int \frac{d^2 {\bf q}_{2}}{(2\pi)^2}
e^{-i{\bf q}_{2}\cdot{\bf b}_{2}} v(0,{\bf q}_{2}) \; \times \nonumber  \\[1ex]
&\;& \times 2ig_s\, \frac{{\bf \epsilon}
\cdot({\bf k}-{\bf q}_{1})}
{({\bf k} - {\bf q}_{1})^2+M^{2}x^{2}+m^{2}_{g}} 
\; \{ e^{i( \frac{ \mathbf{k}^{2} + m_{g}^{2}+M^{2}x^{2}}{2\omega }-q_{2z})
(z_{1}-z_{0})} -  e^{i \frac{ \mathbf{k}^{2} - \mathbf{(k} - 
\mathbf{q}_{1})^{2}}{2\omega} (z_{1}-z_{0})} \}
 \; a_2 [c,a_1] (T_{a_2}T_{a_1})\;\;.
\eeqar{201recov}

In the massless limit this equation reduces to Eq.~(D6) from~\cite{GLV}.

Notice that, unlike to the previous examples, there is no factor of 
$\frac{1}{2}$ in $M_{2,0,1}^{c}$.

We can get $M_{2,0,2}^{c}$ from $M_{2,0,1}^{c}$ by replacing every 2 with 1
and \textit{vice verse}.

However (since scattering centers are identical), we need to symmetrize this
two diagrams, which effectively leads to multiplying every diagram with 
$\frac{1}{2}$.

\section{Zero measure contact limit of $M_{2,1,0}$ and $M_{2,1,1}$}

In calculating the different contributions coming from two interactions with 
the same potential centered around $\vec{\bf x}_1$ we have to  take into
account the two graphs given in Fig.~12, where one of the hits occurs before
the gluon emission vertex and the other one after. 

\begin{center}
\vspace*{6.cm}
\includegraphics{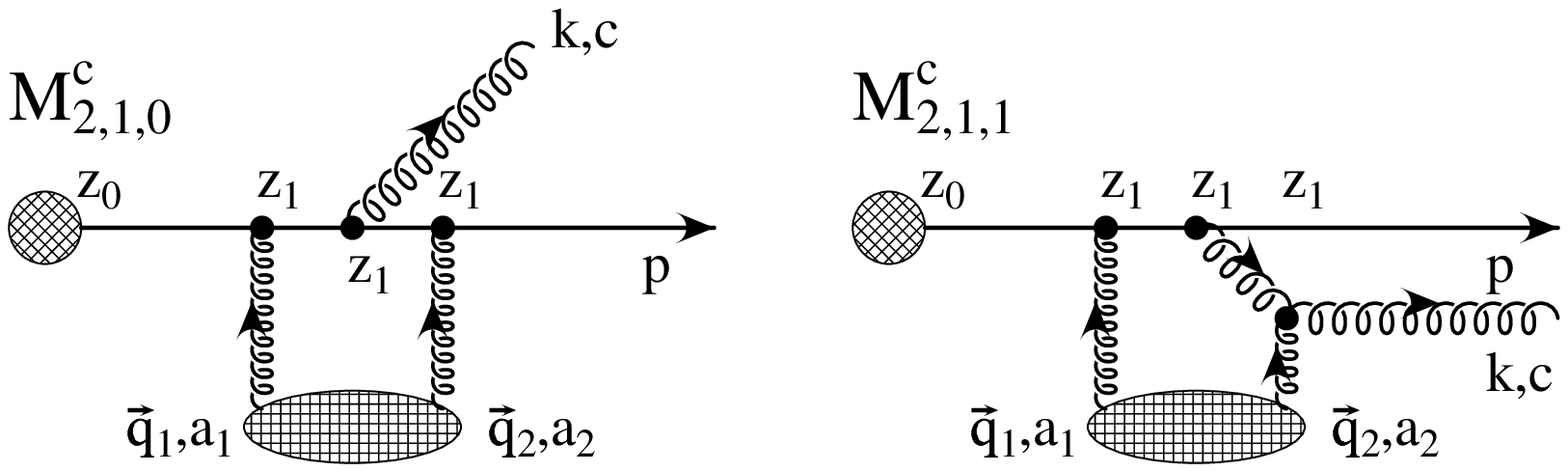}  
\vskip -20pt
\begin{minipage}[t]{15.0cm}
{\small { FIG~12.}
Diagrams  $M^c_{2,1,0}$  and $M^c_{2,1,1}$ of ${\cal O}(0)$ 
according to the time-ordered perturbation theory.} 
\end{minipage}
\end{center}

In the framework of time-ordered perturbation theory 
of~\cite{GLV-Yukawa,GLV1B} the graphs are identically zero
in the contact limit because $\int_{t_1}^{t_1} dt \cdots \equiv 0$. Here we 
present a more detailed study of the validity of this argument in the case of 
massive quarks and gluons.

\beqar
M_{2,1,0} &=& \int \frac{ d^{4} q_{1}}{(2\pi)^{4}} 
\frac{d^{4}q_{2}}{(2\pi)^{4}} i J(p+k-q_{1}-q_{2}) e^{i(p+k-q_{1}-q_{2})x_{0}} \; \times \nonumber  \\[1ex]
&\;& \times \; 
i \Delta_{M} (p+k-q_{1}-q_{2}) V(q_{1}) e^{iq_{1}x_{1}} i \Delta_{M} (p+k-q_{2}) (2p-2q_{2}+k)^{\mu} \epsilon_{\mu } i \Delta_{M} (p-q_{2}) 
ig_{s} V(q_{2}) e^{iq_{2}x_{2}} \; \times \nonumber  \\[1ex]
&\;& \times \;  
(-i (2p+2k-q_{1}-2q_{2})^{0}) (-i (2p-q_{2})^{0}) a_{2} c a_{1} 
T_{a_{2}} T_{a_{1}} \;
\nonumber \\[1ex] 
&\approx& J(p+k) e^{i(p+k)x_{0}} a_{2}ca_{1} (T_{a_{2}} 
T_{a_{1}}) (-ig_{s}) (2E)^{2} \; \times \nonumber  \\[1ex]
&\;& \times \; 
\int \frac{d^{3} \vec{\mathbf{q}}_{1}}{(2\pi)^{3}}
\frac{d^{3} \vec{\mathbf{q}}_{2}}{(2\pi)^{3}} v( \vec{ \mathbf{q}}_{1}) 
e^{-i \vec{ \mathbf{q}}_{1} ( \vec{ \mathbf{x}}_{1} - \vec{\mathbf{x}}_{0})}
v( \vec{ \mathbf{q}}_{2}) e^{-i \vec{\mathbf{q}}_{2}
( \vec{ \mathbf{x}}_{2} - \vec{\mathbf{x}}_{0})} (2p-2q_{2}+k)^{\mu}
\epsilon_{\mu } \; \times \nonumber  \\[1ex]
&\;& \times \; 
\Delta_{M} (p+k-q_{1}-q_{2}) \Delta_{M} (p+k-q_{2})
\Delta_{M} (p-q_{2}) \;
\nonumber \\[1ex] 
&=& J(p+k) e^{i(p+k)x_{0}} a_{2}ca_{1} (T_{a_{2}} T_{a_{1}}) (-ig_{s}) 
(2E)^{2} \int \frac{d^{2} \mathbf{q}_{1}}{(2\pi )^{2}}
\frac{ d^{2} \mathbf{q}_{2}}{(2\pi)^{2}}
e^{ -i \mathbf{q}_{1} \mathbf{b}_{1}} e^{-i \mathbf{q}_{2} \mathbf{b}_{2}}  \; \times \nonumber  \\[1ex]
&\;& \times \; 
\int \frac{dq_{2z}}{2\pi} (2p-2q_{2}+k)^{\mu} \epsilon_{\mu}
\frac{ v( \vec{ \mathbf{q}}_{2}) e^{-iq_{2z}(z_{2}-z_{1})}}
{((p-q_{2})^{2} - M^{2} + i\epsilon) ((p+k-q_{2})^{2} - M^{2} + i\epsilon )}
\int \frac{dq_{z}}{2\pi}
\frac{v( q_{z} - q_{2z}, \mathbf{q}_{1}) e^{-iq_{z}(z_{1}-z_{0})}}
{((p+k-q)^{2} - M^{2}+i\epsilon)}\;
\nonumber \\[1ex] 
&=& J(p+k) e^{i(p+k)x_{0}} a_{2}ca_{1} (T_{a_{2}} T_{a_{1}}) (-ig_{s}) (2E)^{2}
\int \frac{ d^{2} \mathbf{q}_{1}}{(2\pi)^{2}} 
\frac{ d^{2} \mathbf{q}_{2}}{(2\pi)^{2}}
e^{ -i \mathbf{q}_{1} \mathbf{b}_{1}} e^{ -i \mathbf{q}_{2} \mathbf{b}_{2}}
\frac{ \mathbf{ \epsilon \cdot k}}{x} \; \times \nonumber  \\[1ex]
&\;& \times \; 
\int \frac{ dq_{2z}}{2\pi} 
\frac{ v( \vec{ \mathbf{q}}_{2}) e^{ -iq_{2z}(z_{2}-z_{1})}}
{((p-q_{2})^{2} - M^{2} + i\epsilon) ((p+k-q_{2})^{2} - M^{2} + i\epsilon)}  \; \times \nonumber  \\[1ex]
&\;& \times \; 
\frac{-i}{2p_{z}}
v( \frac{ \mathbf{k}^{2} + m_{g}^{2}+M^{2}x^{2}}{2\omega} - q_{2z},
\mathbf{q}_{1}) e^{i \frac{ \mathbf{k}^{2} + m_{g}^{2}+M^{2}x^{2}}{2\omega}
(z_{1}-z_{0})}
\eeqar{210}

In the contact limit, there are four roots which contribute to the integral 
over $q_{2z}$ ($-i\mu_{2}$, $ -i\mu_{1}$, 
$ \frac{ \mathbf{q}_{2}^{2}}{2p_{z}}$ and 
$ ( -\omega _{0} - \tilde{\omega}_{m} )$), leading to

\beqar
M_{2,1,0}^{c}&=&J(p+k) e^{i(p+k)x_{0}} a_{2}ca_{1} (T_{a_{2}}T_{a_{1}})(ig_{s})
\int \frac{ d^{2} \mathbf{q}_{1}}{(2\pi)^{2}}
\frac{ d^{2} \mathbf{q}_{2}}{ (2\pi)^{2}}
e^{ -i( \mathbf{q}_{1} + \mathbf{q}_{2}) \mathbf{b}_{1}} \; \times \nonumber  \\[1ex]
&\;& \times \; 
v(0, \mathbf{q}_{2}) v(0, \mathbf{q}_{1})
\frac{( \mathbf{ \epsilon \cdot k)}}{\mathbf{k}^{2} + m_{g}^{2}+M^{2}x^{2}}
e^{i\frac{ \mathbf{k}^{2} + m_{g}^{2}+M^{2}x^{2}}{2\omega } (z_{1}-z_{0})}
\frac{1}{2i} \frac{ \mathbf{k}^{2} + m_{g}^{2}+M^{2}x^{2}}
{2\omega } \frac{\mu_{1}^{2} + \mu_{1} \mu_{2} + \mu_{2}^{2}} 
{\mu_{1} \mu _{2}( \mu_{1} + \mu_{2})} \;
\nonumber \\[1ex] 
&\sim&
\frac{ \omega_{0}}{\mu} J(p+k) e^{i(p+k)x_{0}} a_{2}ca_{1} 
(T_{a_{2}} T_{a_{1}}) (ig_{s})\; \times \nonumber  \\[1ex]
&\;& \times \; 
\int \frac{ d^{2} \mathbf{q}_{1}}{(2\pi)^{2}} 
\frac{ d^{2} \mathbf{q}_{2}}{(2\pi)^{2}} 
e^{-i( \mathbf{q}_{1} + \mathbf{q}_{2}) \mathbf{b}_{1}}
v(0, \mathbf{q}_{2}) v(0, \mathbf{q}_{1}) 
\frac{( \mathbf{ \epsilon \cdot k)}}{ \mathbf{k}^{2} + m_{g}^{2}+M^{2}x^{2}}
e^{ i \frac{ \mathbf{k}^{2} + m_{g}^{2}+M^{2}x^{2}}{2\omega }(z_{1}-z_{0})}
\eeqar{210fin}

Therefore, the contribution of  $M_{2,1,0}^{c}$ is suppressed by an 
$O(\frac{\omega_{0}}{\mu})$ factor relative to the other graphs.

\bigskip 

\beqar
M_{2,1,1} &=& \int \frac{d^{4}q_{1}}{(2\pi)^{4}} 
\frac{d^{4}q_{2}}{(2\pi)^{4}} iJ(p+k-q_{1}-q_{2}) e^{i(p+k-q_{1}-q_{2}) x_{0}}
\; \times \nonumber  \\[1ex]
&\;& \times \; 
i \Delta_{M} (p+k-q_{1}-q_{2}) (-i (2p+2k-q_{1}-2q_{2})^{0}) V(q_{1})
e^{iq_{1}x_{1}} i \Delta_{M} (p+k-q_{2}) (-i \Delta_{m_{g}} (k-q_{2})) \; \times \nonumber  \\[1ex]
&\;& \times \; 
V(q_{2}) e^{iq_{2}x_{2}}
(-2g_{s}) (2E) (\mathbf{ \epsilon \cdot (k} - \mathbf{q}_{2}))
[c,a_{2}]a_{1} T_{a_{2}} T_{a_{1}} \;
\nonumber \\[1ex] 
&=& 
J(p+k) e^{i(p+k)x_{0}} (-2ig_{s}) (2E)^{2} [c,a_{2}]a_{1} T_{a_{2}}T_{a_{1}}
\; \times \nonumber  \\[1ex]
&\;& \times \;
\int \frac{ d^{3} \vec{ \mathbf{q}}_{1}}{(2\pi)^{3}}
\frac{ d^{3} \vec{ \mathbf{q}}_{2}}{(2\pi)^{3}} v( \vec{ \mathbf{q}}_{1}) 
e^{ -i \vec{ \mathbf{q}}_{1} ( \vec{ \mathbf{x}}_{1} - \vec{\mathbf{x}}_{0})}
v( \vec{ \mathbf{q}}_{2}) e^{ -i \vec{ \mathbf{q}}_{2}
(\vec{ \mathbf{x}}_{2} - \vec{\mathbf{x}}_{0})}
( \mathbf{ \epsilon \cdot (k} - \mathbf{q}_{2}))\; \times \nonumber  \\[1ex]
&\;& \times \;
\Delta_{M} (p+k-q_{1}-q_{2}) \Delta_{M} (p+k-q_{2}) 
\Delta_{m_{g}} (k-q_{2}) \;
\nonumber \\[1ex] 
&=& 
J(p+k) e^{i(p+k)x_{0}} (-2ig_{s}) (2E)^{2} [c,a_{2}]a_{1} T_{a_{2}}T_{a_{1}}
\int \frac{d^{2}\mathbf{q}_{1}}{(2\pi)^{2}}
\frac{ d^{2} \mathbf{q}_{2}}{(2\pi)^{2}}
e^{ -i \mathbf{q}_{1} \mathbf{b}_{1}} 
e^{ -i \mathbf{q}_{2} \mathbf{b}_{2}} 
( \mathbf{ \epsilon \cdot (k} - \mathbf{q}_{2})) \; \times \nonumber  \\[1ex]
&\;& \times \;
\int \frac{ dq_{2z}}{2\pi } \frac{ v( \vec{ \mathbf{q}}_{2})
e^{ -iq_{2z} (z_{2}-z_{1})}}{((k-q_{2})^{2} - m_{g}^{2} + i\epsilon )
((p+k-q_{2})^{2} - M^{2} + i\epsilon )} 
\int \frac{dq_{z}}{2\pi} \frac{v(q_{z}-q_{2z}, \mathbf{q}_{1})
e^{ -iq_{z} (z_{1}-z_{0})}} {((p+k-q)^{2} - M^{2} + i\epsilon)} \;
\nonumber \\[1ex] 
&=& J(p+k) e^{i(p+k)x_{0}} (-2ig_{s}) (2E)^{2} [c,a_{2}]a_{1} T_{a_{2}}T_{a_{1}}\; \int \frac{ d^{2} \mathbf{q}_{1}}{(2\pi)^{2}}
\frac{ d^{2} \mathbf{q}_{2}}{(2\pi)^{2}} 
e^{-i \mathbf{q}_{1} \mathbf{b}_{1}} e^{-i \mathbf{q}_{2} \mathbf{b}_{2}}
(\mathbf{ \epsilon \cdot (k} - \mathbf{q}_{2})) \; \times \nonumber  \\[1ex]
&\;& \times \;
\int \frac{dq_{2z}}{2\pi } \frac{v( \vec{ \mathbf{q}}_{2})
e^{ -iq_{2z} (z_{2}-z_{1})}} {((k-q_{2})^{2} - m_{g}^{2} + i\epsilon)
((p+k-q_{2})^{2} - M^{2} + i\epsilon )}\; \times \nonumber  \\[1ex]
&\;& \times \;
\frac{-i}{2p_{z}} v( \frac{ \mathbf{k}^{2} + m_{g}^{2}+M^{2}x^{2}}
{2\omega} - q_{2z}, \mathbf{q}_{1}) 
e^{ i \frac{ \mathbf{k}^{2} + m_{g}^{2}+M^{2}x^{2}}{2\omega } (z_{1}-z_{0})}
\eeqar{211}

In the contact limit, there are also four roots which contribute to the 
integral over $q_{2z}$ ($-i\mu_{2}$, $ -i\mu_{1}$, 
$ \frac{ \mathbf{q}_{2}^{2}}{2p_{z}}$ 
and $ (\omega_{2} - \omega _{0})$). After computing the residues, we get

\beqar
M_{2,1,1}^{c} &=& J(p+k) e^{i(p+k)x_{0}} (2ig_{s}) [c,a_{2}]a_{1} 
T_{a_{2}} T_{a_{1}} \; \times \nonumber  \\[1ex]
&\;& \times \; 
\int \frac{d^{2}\mathbf{q}_{1}}{(2\pi)^{2}}
\frac{ d^{2} \mathbf{q}_{2}}{(2\pi)^{2}} 
e^{ -i (\mathbf{q}_{1} + \mathbf{q}_{2}) \mathbf{b}_{1}}
v(0, \mathbf{q}_{2}) v(0, \mathbf{q}_{1}) 
\frac{( \mathbf{ \epsilon \cdot (k} - \mathbf{q}_{1})}
{ \mathbf{(k} - \mathbf{q}_{1})^{2} + m_{g}^{2}+M^{2}x^{2}}  \; \times \nonumber  \\[1ex]
&\;& \times \; 
e^{ i \frac{\mathbf{k}^{2} + M^{2}x^{2 } + m_{g}^{2}}{2\omega }
(z_{1}-z_{0})} \frac{1}{2i} 
\frac{\mathbf{(k} - \mathbf{q}_{1})^{2} + m_{g}^{2}+M^{2}x^{2}}{2\omega }
\frac{\mu_{1}^{2} + \mu_{1} \mu_{2} + \mu_{2}^{2}}
{\mu_{1} \mu_{2}( \mu_{1} + \mu_{2})} \;\nonumber  \\[1ex]
&\sim& 
\frac{ \omega_{0}}{\mu} J(p+k) e^{i(p+k)x_{0}} (2ig_{s}) 
[c,a_{2}]a_{1} T_{a_{2}} T_{a_{1}} \; \times \nonumber  \\[1ex]
&\;& \times \; 
\int \frac{ d^{2} \mathbf{q}_{1}}{(2\pi)^{2}}
\frac{ d^{2} \mathbf{q}_{2}}{(2\pi)^{2}} 
e^{-i( \mathbf{q}_{1} + \mathbf{q}_{2}) \mathbf{b}_{1}} v(0,\mathbf{q}_{2})
v(0, \mathbf{q}_{1}) 
\frac{( \mathbf{ \epsilon \cdot k)}}{\mathbf{k}^{2} + m_{g}^{2}+M^{2}x^{2}}
e^{ i \frac{ \mathbf{k}^{2} + m_{g}^{2}+M^{2}x^{2}}{2\omega }(z_{1}-z_{0})}
\eeqar{211fin}

We see that the contribution of $M_{2,1,0}^{c}$ is also suppressed by an 
$ O( \frac{ \omega_{0}}{\mu})$ factor relative to the other graphs. Therefore,
we will neglect the contribution of $M_{2,1,1}^{c}$ and $M_{2,1,0}^{c}$ 
in the energy loss calculation.

\section{Computation of the first order radiative energy loss}

In this section we want to compute the first order in opacity radiative energy loss. According to Eq.~(\ref{FO1}) we have

\beq
d^{3} N_{g}^{(1)} d^{3} N_{J} = \frac{d}{d_{T}}
({\rm Tr} \left\langle |M_{1}|^{2} \right\rangle + 
\frac{2}{d_{T}}{\rm Re}{\rm Tr} \left\langle M_{0}^{\ast }M_{2}\right\rangle ) 
\frac{d^{3}\vec{\mathbf{p}}} {(2\pi )^{3}2p^{0}}
\frac{d^{3}\vec{\mathbf{k}}} {( 2\pi)^{3} 2\omega },
\eeq

$M_{1}$ is sum of all diagrams with one scattering center and $M_{2}$ is sum
of all diagrams with two scattering centers.

Using the results for $M_{1,0,0}$, $M_{1,1,0}$ and $M_{1,0,1}$ from Appendix 
A we get

\beqar
M_{1} &=& M_{1,0,0} + M_{1,1,0} + M_{1,0,1} = 
J(p) e^{i(p+k)x_{0}} (-i) (2ig_{s}) T_{a_{1}}
\int \frac{d^{2}\mathbf{q}_{1}}{( 2\pi ) ^{2}}
v(0,\mathbf{q}_{1}) e^{-i\mathbf{q}_{1}\mathbf{b}_{1}}\nonumber \\[1ex]
&\times& \{ ( \frac{(\mathbf{\epsilon \cdot (k}-\mathbf{q}_{1}))}
{(\mathbf{k-q}_{1})^{2} + m_{g}^{2}+M^{2}x^{2}} -  
\frac{\mathbf{\epsilon \cdot k}}{\mathbf{k}^{2} + m_{g}^{2}+M^{2}x^{2}})
e^{i(\omega _{0} + \tilde{\omega}_{m}) (z_{1}-z_{0})} [c,a_{1}] -\nonumber \\[1ex]
&-& \frac{(\mathbf{\epsilon \cdot (k}-\mathbf{q}_{1}))} 
{(\mathbf{k-q}_{1})^{2} + m_{g}^{2}+M^{2}x^{2}}
e^{i(\omega_{0}-\omega_{1})(z_{1}-z_{0})} [c,a_{1}] -
\frac{\mathbf{\epsilon \cdot k}} {\mathbf{k}^{2} + m_{g}^{2}+M^{2}x^{2}}
a_{1}c \},
\eeqar{}

where

$\omega_{1} = \frac{(\mathbf{k-q}_{1})^{2}+m_{g}^{2}} { 2 \omega }$, and 
$\tilde{\omega}_{m} = \frac{M^{2}x^{2}}{2\omega}$.

\medskip

Then,

\beqar
\frac{1}{d_{T}} \left\langle |M_{1}|^{2}\right\rangle &=& 
N |J(p)|^{2} (4g_{s}^{2}) \frac{1}{A_{\perp }} 
\int \frac{d^{2}\mathbf{q}_{1}} {(2\pi)^{2}}
|v(\mathbf{q}_{1})|^{2} \frac{C_{2}(T)}{d_{A}} \times \nonumber \\[1ex] 
&\times& \{ 2 \alpha ( \frac{(\mathbf{\epsilon \cdot (k}-\mathbf{q}_{1}))}
{(\mathbf{k-q}_{1})^{2} + m_{g}^{2}+M^{2}x^{2}} - 
\frac{\mathbf{\epsilon \cdot k}}
{\mathbf{k}^{2} + m_{g}^{2}+M^{2}x^{2}})^{2} + 
2\alpha ( \frac{(\mathbf{\epsilon \cdot (k}-\mathbf{q}_{1}))}
{(\mathbf{k-q}_{1})^{2} + m_{g}^{2}+M^{2}x^{2}})^{2} - \nonumber \\[1ex]
&-&2 \alpha ( \frac{(\mathbf{\epsilon \cdot (k} - \mathbf{q}_{1}))}
{(\mathbf{k-q}_{1})^{2} + m_{g}^{2}+M^{2}x^{2}} -
\frac{\mathbf{ \epsilon \cdot k}} {\mathbf{k}^{2}+m_{g}^{2}+M^{2}x^{2}})
\frac{(\mathbf{\epsilon \cdot (k} - \mathbf{q}_{1}))}
{(\mathbf{k-q}_{1})^{2} + m_{g}^{2}+M^{2}x^{2}}
2 \cos ((\omega_{1}+\tilde{\omega}_{m})(z_{1}-z_{0})) -\nonumber \\[1ex] 
&-& \alpha ( \frac{(\mathbf{\epsilon \cdot (k} - \mathbf{q}_{1}))}
{(\mathbf{k-q}_{1})^{2} + m_{g}^{2}+M^{2}x^{2}} - 
\frac{\mathbf{ \epsilon \cdot k}}{\mathbf{k}^{2} + m_{g}^{2}+M^{2}x^{2}})
\frac{\mathbf{ \epsilon \cdot k}}{\mathbf{k}^{2} + m_{g}^{2}+M^{2}x^{2}}
2 \cos ((\omega_{0} + \tilde{\omega}_{m})(z_{1}-z_{0})) +\nonumber \\[1ex]
&+& \alpha \frac{\mathbf{\epsilon \cdot k}}
{\mathbf{k}^{2} + m_{g}^{2}+M^{2}x^{2}}
\frac{(\mathbf{ \epsilon \cdot (k} - \mathbf{q}_{1}))}
{(\mathbf{k-q}_{1})^{2} + m_{g}^{2}+M^{2}x^{2}}
2 \cos ((\omega_{0} - \omega _{1})(z_{1}-z_{0}))  +\nonumber \\[1ex] 
&+&(\frac{ \mathbf{ \epsilon \cdot k}}{\mathbf{k}^{2}+m_{g}^{2}+M^{2}x^{2}})^{2}Tr a^{2} c^{2}\}
\eeqar{}

Here we used $ {\rm Tr}(T_{a_{1}}T_{a_{2}}) = \frac{C_{2}(T)d_{T}}{d_{A}} 
\delta_{_{a_{1}a_{2}}} $ , and defined $ \alpha \equiv Tr(c^{2}a^{2}-caca) $. 
Factor $N$ comes from sum over $N$ scattering centers.

To compute $M_{2}$ we will first add all the diagrams with two scattering 
centers calculated in Appendix B-E, and than take their average.

\beqar
M_{2} &=& \frac{1}{2} J(p) e^{i(p+k)x_{0}} (-2ig_{s}) T_{a_{1}} T_{a_{2}}
\frac{1}{A_{\perp}} \int \frac{d^{2}\mathbf{q}_{1}}{(2\pi)^{2}}
| v( \mathbf{q}_{1} ) |^{2} \nonumber \\[1ex]
&\times& \{ \frac{\mathbf{\epsilon \cdot k}}
{\mathbf{k}^{2} + m_{g}^{2}+M^{2}x^{2}} 
\{ e^{i( \omega_{0} + \tilde{\omega}_{m}) (z_{1}-z_{0})}
([[c,a_{2}],a_{1}] + [a_{2}a_{1},c])-[[c,a_{2}],a_{1}] - a_{2}a_{1}c \} - \nonumber \\[1ex]
&-& \frac{(\mathbf{ \epsilon \cdot (k} - \mathbf{q}_{1}))}
{(\mathbf{k-q}_{1})^{2} + m_{g}^{2}+M^{2}x^{2}} 
(e^{ i ( \omega_{0} + \tilde{\omega}_{m})(z_{1}-z_{0})} -
e^{ i ( \omega_{0} - \omega_{1} )(z_{1}-z_{0})}) 
(a_{2}[c,a_{1}] + a_{1}[c,a_{2}] ) \}
\eeqar{}

Using this we can now find $ \frac{2}{d_{T}} {\rm Re} {\rm Tr} 
\left\langle M_{0}^{\ast} M_{2} \right\rangle$.

\beqar
\frac{2}{d_{T}} {\rm Re} {\rm Tr} 
\left\langle M_{0}^{\ast } M_{2} \right\rangle
&=& N |J(p)|^{2} (4g_{s}^{2}) \frac{1}{A_{\perp }} 
\int \frac{d^{2}\mathbf{q}_{1}}{( 2\pi)^{2}}| v( \mathbf{q}_{1}) |^{2}
\frac{C_{2}(T)}{d_{A}}\nonumber \\[1ex]
&\times& \{ ( \frac{ \mathbf{ \epsilon \cdot k }}
{ \mathbf{k}^{2} + m_{g}^{2}+M^{2}x^{2}})^{2}
(2 \alpha \cos ((\omega_{0} + \omega^{\prime })(z_{1}-z_{0})) - 
2 \alpha - Tra^{2} c^{2}) + \nonumber \\[1ex]
&+& 2 \alpha \frac{ \mathbf{ \epsilon \cdot k}} 
{\mathbf{k}^{2} + m_{g}^{2}+M^{2}x^{2}}
\frac{(\mathbf{ \epsilon \cdot (k} - \mathbf{q}_{1}))}
{ (\mathbf{k-q}_{1})^{2} + m_{g}^{2}+M^{2}x^{2}}\nonumber \\[1ex]
&\times& \{ \cos ((\omega_{0} + \tilde{\omega}_{m})(z_{1}-z_{0})) - 
\cos ((\omega_{0} - \omega_{1})(z_{1}-z_{0})) \} \}
\eeqar{}

Therefore, $ \frac{1}{d_{T}} \left\langle |M_{1}|^{2} \right\rangle + 
\frac{2}{d_{T}} {\rm Re} {\rm Tr} \left\langle M_{0}^{\ast} M_{2} \right
\rangle$ is equal to

\beqar
\frac{1}{d_{T}} \left\langle |M_{1}|^{2} \right\rangle &+& 
\frac{2}{d_{T}} {\rm Re} {\rm Tr} \left\langle M_{0}^{\ast } M_{2} 
\right\rangle = D_{R} |J(p)|^{2} (4g_{s}^{2}) \frac{C_{2}(T)}{d_{A}}C_{R}^{2} 
\frac{1}{A_{\perp}} \int \frac{d^{2}\mathbf{q}_{1}}{(2\pi )^{2}}
| v( \mathbf{q}_{1}) | ^{2}\nonumber \\[1ex]
&\times& \{ -2 \frac{(\mathbf{\epsilon \cdot (k} - \mathbf{q}_{1}))}
{ (\mathbf{k-q}_{1})^{2} + m_{g}^{2}+M^{2}x^{2}}
( \frac{\mathbf{\epsilon \cdot k}}{\mathbf{k}^{2} + m_{g}^{2}+M^{2}x^{2}} - 
\frac{ ( \mathbf{ \epsilon \cdot (k} - \mathbf{q}_{1}))}
{(\mathbf{k-q}_{1})^{2} + m_{g}^{2}+M^{2}x^{2}}) \times \nonumber \\[1ex]
&\times& (1 - \cos ( \frac{( \mathbf{k}-\mathbf{q}_{1})^{2} +m_{g}^{2}+
M^{2}x^{2}}{2p_{z}x}(z_{1}-z_{0}))) \}.
\eeqar{}

Here we have used that $\omega_{1} + \tilde{\omega}_{m} = 
\frac{( \mathbf{k} - \mathbf{q}_{1})^{2} + m_{g}^{2}+M^{2}x^{2}}{2p_{z}x}$ and 
$\alpha=\frac{1}{2} C_{R}^{2} D_{R} $.

\medskip

Using Eqs. (\ref{FO1},~\ref{FO2}) it is now easy to extract $d E_{ind}^{(1)}
\equiv \omega d^{3} N_{g}$

\beqar
\frac {d E_{ind}^{(1)}}{dx} &=& \frac{C_{R} \alpha_{S}}{\pi} \frac{L}{\lambda} 
E \int \frac{d^{2} \mathbf{q}_{1}} {\pi} 
\frac{\mu^{2}}{(\mathbf{q}_{1}^2 +\mu^{2})^{2}} \int 
\frac{d^{2}\mathbf{k}}{\pi}  \times \nonumber \\[1ex]
&\times& \{ -2 \frac{(\mathbf{\epsilon \cdot (k} - \mathbf{q}_{1}))}
{ (\mathbf{k-q}_{1})^{2} + m_{g}^{2}+M^{2}x^{2}}
( \frac{\mathbf{\epsilon \cdot k}}{\mathbf{k}^{2} + m_{g}^{2}+M^{2}x^{2}} - 
\frac{ ( \mathbf{ \epsilon \cdot (k} - \mathbf{q}_{1}))}
{(\mathbf{k-q}_{1})^{2} + m_{g}^{2}+M^{2}x^{2}}) \times \nonumber \\[1ex]
&\times& \int dz_{1} (1 - \cos ( \frac{( \mathbf{k}-\mathbf{q}_{1})^{2} +
m_{g}^{2}+ M^{2}x^{2}}{2p_{z}x}(z_{1}-z_{0}))) \}
\frac{ e^{ - \frac{(z_{1}-z_{0})}{L}}}{L}\; \; ,
\eeqar{dE1}

where we have used Eqs.~(1, 6) from~\cite{GLV} to write the result in terms 
of opacity $L/\lambda$. For simplicity we have also assumed exponential 
distribution $\exp(-\Delta z/L)/L$ between scattering centers. After the $z_{1}$ integration we get

\beqar
\frac{ d E_{ind}^{(1)}}{d x} &=& \frac{C_{R} \alpha_{S}}{\pi} 
\frac{L}{\lambda} E \int \frac {d \mathbf{k}^{2}}{\mathbf{k}^{2} + 
m_{g}^{2}+M^{2}x^{2}} \int \frac{d^{2} \mathbf{q}_{1}} {\pi} 
\frac{\mu^{2}}{(\mathbf{q}_{1}^2 +\mu^{2})^{2}} \times \nonumber \\[1ex]
&\times& 2 \; \; \frac{ \mathbf{k} \cdot \mathbf{q}_{1} 
( \mathbf{k}-\mathbf{q}_{1})^{2} + (m_{g}^{2}+M^{2}x^{2})\mathbf{q}_{1} \cdot 
(\mathbf{q}_{1}-\mathbf{k})}{( \frac{4 E x}{L} )^{2} 
+ (( \mathbf{k}-\mathbf{q}_{1})^{2} + M^{2}x^{2} + m_{g}^{2})^{2}}
\eeqar{dE/dx}

which in the massless limit reduces to Eq~(125) from~\cite{GLV}.
\end{appendix}

\end{document}